\documentclass[5p,times]{elsarticle}
\usepackage{ecrc}
\volume{00}
\firstpage{1}

\CopyrightLine{2025}{Published by Elsevier Ltd.}

\usepackage[figuresright]{rotating}
\usepackage{color}
\usepackage{subfigure}
\usepackage{multirow} 
\usepackage{xcolor}
\usepackage{url}

\usepackage{mathrsfs}
\usepackage{amsmath}
\usepackage{amssymb}
\usepackage{setspace}
\usepackage{epstopdf}
\usepackage{booktabs}
\usepackage{array}
\usepackage{siunitx,etoolbox} 

\usepackage{makecell}
\usepackage{enumitem} 
\usepackage{tabularx}

\usepackage{algorithm}
\usepackage{algpseudocode}
\algnewcommand{\Input}[1]{\Statex\textbf{Input:} #1}
\algnewcommand{\Parameter}[1]{\Statex\textbf{Parameter:} #1}
\algnewcommand{\Output}[1]{\Statex\textbf{Output:} #1}

\newtheorem{theorem}{Theorem}

\newtheorem{Proposition}{Proposition}
\newtheorem{definition}{Definition}
\newtheorem{lemma}{Lemma}
\newtheorem{remark}{Remark}
\newtheorem{assumption}{Assumption}

\begin{document}

\begin{frontmatter}


\title{\Large \bf Learning-based data-enabled economic predictive control with convex optimization for nonlinear systems \tnoteref{abc}}

\tnotetext[abc]{This research is supported by the National Research Foundation, Singapore, and PUB, Singapore’s National Water Agency under its RIE2025 Urban Solutions and Sustainability (USS) (Water) Centre of Excellence (CoE) Programme, awarded to Nanyang Environment \& Water Research Institute (NEWRI), Nanyang Technological University, Singapore (NTU). This research is also supported by the Ministry of Education, Singapore, under its Academic Research Fund Tier 1 (RG95/24 and RG63/22), and Nanyang Technological University, Singapore (Start-Up Grant). Any opinions, findings and conclusions or recommendations expressed in this material are those of the author(s) and do not reflect the views of the National Research Foundation, Singapore and PUB, Singapore's National Water Agency. 
}

\author[NEWRI,IGP,CCEB]{Mingxue Yan}
\author[CCEB]{Xuewen Zhang}
\author[ME]{Kaixiang Zhang}
\author[ME]{Zhaojian Li}
\author[NEWRI,CCEB]{Xunyuan Yin\corref{cor1}}
\address[NEWRI]{Nanyang Environment and Water Research Institute, Nanyang Technological University, Singapore}
\address[IGP]{Interdisciplinary Graduate Programme, Nanyang Technological University, Singapore}
\address[CCEB]{School of Chemistry, Chemical Engineering and Biotechnology, Nanyang Technological University, Singapore}
\address[ME]{Department of Mechanical Engineering, Michigan State University, USA}
\cortext[cor1]{Corresponding author: X. Yin. Email: xunyuan.yin@ntu.edu.sg.}

\begin{abstract}
In this article, we propose a data-enabled economic predictive control method for a class of nonlinear systems, which aims to optimize the economic operational performance while handling hard constraints on the system outputs. Two lifting functions are constructed via training neural networks, which generate mapped input and mapped output in a higher-dimensional space, where the nonlinear economic cost function can be approximated using a quadratic function of the mapped variables. 
The data-enabled predictive control framework is extended to address nonlinear dynamics by using the mapped input and the mapped output that belong to a virtual linear representation, which serves as an approximation of the original nonlinear system.
Additionally, we reconstruct the system output variables from the mapped output, on which hard output constraints are imposed. The online control problem is formulated as a convex optimization problem, despite the nonlinearity of the system dynamics and the original economic cost function.
Theoretical analysis is presented to justify the suitability of the proposed method for nonlinear systems.  
We evaluate the proposed method through two large-scale industrial case studies: (i) a biological water treatment process, and (ii) a solvent-based shipboard post-combustion carbon capture process. These studies demonstrate its effectiveness and advantages.
\end{abstract}


\end{frontmatter}

\section{Introduction}
Modern industrial processes exhibit characteristics such as highly nonlinear dynamics and susceptibility to unknown disturbances, which pose challenges to system control. Conventional linear control methods are difficult to address nonlinear systems, due to their inability to comprehensively capture the nonlinear dynamic behaviors, potentially leading to suboptimal performance or instability~\cite{henson1997nonlinear}. Advanced control approaches are required to regulate the operational performance and guarantee safety. 

In recent years, numerous nonlinear control methods have been proposed to address processes with significant nonlinearity. One representative method is nonlinear model predictive control (MPC), which computes optimal control inputs by minimizing an objective function over a predictive horizon~\cite{schwenzer2021review, rawlings2000tutorial, grune2017nonlinear}. The robustness of nonlinear MPC has been extensively studied since unknown disturbances and noise are prevalent in real-world applications. In~\cite{morari2012nonlinear}, the offset-free MPC for nonlinear systems was proposed by constructing an augmented state containing original system states and disturbances. In~\cite{mayne2011tube}, a tube-based nonlinear MPC scheme was proposed to enhance the robustness against additive disturbances.
A non-trivial extension of nonlinear MPC, economic model predictive control (EMPC), optimizes the economic performance effectively by incorporating economic considerations into the control objective function~\cite{ellis2014tutorial, ellis2017economic}. EMPC offers significant flexibility, as it does not stabilize the state variables at constant levels, but instead drives the system to pursue optimal economic operations, making it more adaptable to complex and dynamic market environments. In~\cite{rawlings2012fundamentals}, EMPC with terminal constraints and terminal cost was introduced, and the stability was analyzed based on dissipativity properties. \cite{faulwasser2018economic} revisited a variety of EMPC designs, including schemes with and without terminal constraints and penalties.
Although MPC is effective for nonlinear systems, the formulation of the control optimization problem depends on an accurate first-principles model, which is usually unavailable in practical applications due to the significant time investment and the requirement of expert knowledge. 

To address these problems, data-driven modeling and control have emerged as promising alternatives. One of the promising data-driven control frameworks is Koopman-based model predictive control~\cite{korda2018linear,arbabi2018data}. According to the Koopman operator theory, the dynamics of a nonlinear system can be described by a surrogate linear Koopman model with the lifted states~\cite{koopman1931hamiltonian}. A linear Koopman model can be constructed using purely data-driven approaches and historical data, thereby addressing the absence of a first-principles model. By replacing the nonlinear dynamics with the established Koopman model, a linear MPC problem can be formulated \cite{korda2018linear}. The computational efficiency is significantly improved compared to the nonlinear MPC scheme. Although Koopman-based MPC addresses the lack of first-principles models, a remaining issue is the requirement of full-state measurements. Since Koopman modeling typically relies on full-state information, it is necessary to integrate Koopman-based state estimation to provide real-time state estimation~\cite{ yan2025self}. Exceptions can be found in~\cite{han2024efficient}. In~\cite{han2024efficient}, a learning-based Koopman model was established using solely input and output data. A convex Koopman-based EMPC is then formulated by approximating the economic cost function using a quadratic term. In~\cite{tang2025big}, a linearization approach similar to the concept of Koopman modeling was employed. An autoencoder was used to project the nonlinear system into a linear latent variable space, where the linear property was enforced through a kernel representation. Behavioral system theory was utilized to treat the control input and system output trajectories jointly in the latent space.

In recent years, the data-enabled predictive control (DeePC) framework has attracted growing attention since it uses non-parametric representations of linear time-invariant (LTI) systems to bypass the need for system modeling and state estimation~\cite{coulson2019data}. Within the DeePC framework, input and output data are used to construct Hankel matrices that describe the linear system dynamics according to Willems' fundamental lemma~\cite{1428856}.
In \cite{huang2023robust}, a robust DeePC approach was proposed by formulating a min-max control optimization problem.
In~\cite{yang2015data, zhang2023dimension}, singular value decomposition (SVD) was employed to reduce the order of Hankel matrices and enhance online computational efficiency. In~\cite{shang2024willems,zhang2025deep}, the DeePC framework was leveraged to handle nonlinear systems. 
In \cite{berberich2022linear}, a data-driven MPC for nonlinear systems with stability guarantee was proposed based on the fundamental lemma, using local linearization of the nonlinear system dynamics. In~\cite{xie2023linear}, a data-driven EMPC framework was proposed to address linear systems. 
In~\cite{li2024physics}, a physics-augmented DeePC was proposed to address system variations and uncertainties and developed an eco-driving optimization framework to reduce the operational energy consumption for connected and automated vehicles. 
In \cite{WR2025}, a mixed-integer economic zone DeePC approach was proposed to extend conventional DeePC from set-point tracking to zone tracking and energy-efficient operation of connected open water systems subject to practical state and input constraints. 

Based on the above observations, in this work, we propose a data-enabled economic predictive control (DeeEPC) scheme to optimize the economic performance of a class of nonlinear systems using the DeePC framework. To preserve convexity, the nonlinear economic cost function is approximated by quadratic functions of a mapped output and a mapped input, which are generated by neural networks trained on open-loop input-output data, thereby enabling efficient online implementation. Hard constraints imposed on the key quality variables are handled by reconstructing the corresponding output variables from the mapped output using a reconstruction matrix. In addition, the mapped output, the mapped input, and the original system input are expected to belong to a virtual linear system that serves as a good approximation of the underlying nonlinear system. It is proven that an infinite-dimensional mapped output exists that can perfectly represent the nonlinear economic cost function while ensuring the feasibility of applying Willems' fundamental lemma~\cite{1428856}. Moreover, the approximation error of the economic cost function is shown to remain bounded when utilizing the finite-dimensional mapped variables.

Partial results of this work have been presented in the conference version~\cite{yan2025economic}. In~\cite{yan2025economic}, we made an initial attempt to develop a DeePC-based model-free economic predictive control scheme for nonlinear systems, where the economic cost function was approximated using a quadratic function of only a transformed output vector obtained from an output-dependent lifting function.
The current work significantly extends the conference version in~\cite{yan2025economic} in several important aspects: 1) As compared to~\cite{yan2025economic}, in the current work, we introduce two lifting functions, one for generating mapped system output and the other for generating mapped control input, which enable appropriate approximations of more general economic cost functions that explicitly depend on both system outputs and control inputs. 2) We conduct a theoretical analysis to justify the suitability of the proposed method for addressing nonlinear processes. 3) We incorporate regularization into the proposed design to handle stochastic disturbances and noise. 4) We demonstrate the effectiveness of the proposed method through the applications to two large-scale industrial processes, including a biological water treatment plant and a shipboard post-combustion carbon capture system, whereas \cite{yan2025economic} considered a chemical process consisting of two reactors.

\section{Preliminaries and problem formulation}
This work aims to propose a data-driven economic predictive control method for general nonlinear systems within the framework of data-enabled predictive control (DeePC) \cite{coulson2019data}, \allowbreak which was originally developed for the control of linear time-invariant (LTI) systems. In order to extend the original DeePC \cite{coulson2019data} to the nonlinear context, we choose to leverage Koopman theory~\cite{korda2018linear,koopman1931hamiltonian}. This section briefly reviews DeePC~\cite{coulson2019data} and Koopman theory~\cite{korda2018linear,koopman1931hamiltonian} as the foundation of the proposed method.

\subsection{Notation}
$\mathbb{N}^+$ denotes the set of positive integers. $\mathbb R$ denotes the set of real numbers. $x = [x_{(1)},\dots,x_{(n)}]^\top$ is a $n$-dimensional column vector, where $x_{(i)}$ denotes the $i$-th element of vector $x$. $\{x\}_i^j$ is a sequence of vector $x$ from time instant $i$ to $j$. $x_{j|k}$ represents the information about vector $x$ for time instant $j$, made available at time instant $k$. $\left\|x\right\|$ represents the Euclidean norm of vector $x$. $\left\|x\right\|^2_P$ is the square of the weighted norm of vector $x$, computed as $\left\|x\right\|^2_P = x^\top P x$. 
$\left\|f(x)\right\|$ represents the $L_2$-norm of function $f(x)$. 
$\langle x, y \rangle $ is the inner product of $x$ and $y$. 
$\mathbf{f}(x) = \{f_i(x)\}_{i = 1}^j$ denotes a sequence of functions $f_i(x)$, $i=1,\ldots,j$. The corresponding vector-valued function is $f(x) = [f_1(x),\dots,f_j(x)]^\top$, obtained by stacking the elements of $\mathbf{f}(x)$.
$\mathbf{f}(x)\bigotimes \mathbf{g}(y) $ represents the product of sequences, computed as $\mathbf{f}(x)\bigotimes \mathbf{g}(y) = \{ f_i(x)g_j(y)  \;\vert \; i\in I, j\in J\}$, where $I$ and $J$ are the set of indices of elements in $\mathbf{f}(x)$ and $\mathbf{g}(y)$, respectively.
$\bigotimes_{n=1}^N \mathbf f_n(x)$ denotes the product of $N$ sequences, computed as $\bigotimes_{n=1}^N\mathbf f_n(x)=\mathbf f_1(x) \bigotimes\mathbf f_2(x)\ldots \bigotimes \mathbf f_N(x)$.
$M \setminus N$ represents the set difference of $M$ and $N$.
$\text{diag}\left( \cdot \right)$ represents a diagonal matrix formed from its arguments. $\text{exp}(\cdot)$ denotes the element-wise exponential operator.
$I_{n}$ is an identity matrix of size $n\times n$. $\bf{0}$ denotes a zero matrix of appropriate size. $\text{det}(A)$ returns the determinant of square matrix $A$.

\subsection{Data-enabled predictive control}
Consider a discrete-time linear time-invariant (LTI) system:
\begin{subequations}\label{ed:linearmodel}
\begin{align}
    x_{k+1} &= Ax_k+Bu_k \label{ed:linearmodel:1}\\
    y_k &= Cx_k+Du_k \label{ed:linearmodel:2}
\end{align}
\end{subequations}
where $x\in\mathbb{X}\subseteq\mathbb{R}^{n_x}$ denotes the state vector; $y\in\mathbb{Y}\subseteq\mathbb{R}^{n_y}$ denotes the system output vector; $u\in\mathbb{U}\subseteq\mathbb{R}^{n_u}$ denotes the control input vector;
$A\in \mathbb{R}^{n_x\times n_x}$, $B\in \mathbb{R}^{n_x\times n_u}$, $C\in \mathbb{R}^{n_y\times n_x}$, and $D\in \mathbb{R}^{n_y\times n_u}$ are system matrices.


According to the Willems' fundamental lemma~\cite{1428856}, with $T$-step historical input and output data collected from system~(\ref{ed:linearmodel}), which are denoted by $\mathbf{u}_{T}^d:= \{u^d \}_1^{T}$ and $\mathbf{y}_{T}^d:= \{ y^d\}_{1}^{T}$, Hankel matrices $\mathscr{H}_L(\mathbf{u}_{T}^d)$ and $\mathscr{H}_L(\mathbf{y}_{T}^d)$ can be constructed as decribed in Definition~\ref{deepc:def:pe}.  

\begin{definition}[Persistent excitation\cite{1428856}]\label{deepc:def:pe}
Let $T, L \in \mathbb{N}^+$ and $T\geq L$. A Hankel matrix of depth $L$ is constructed from the input sequence $\mathbf{u}_T^d$ as follows:
\begin{equation}\label{hankel}
    \mathscr{H}_L(\mathbf{u}_T^d)=\left[\begin{array}{c c c c }
                    u_1^d & u_2^d & \dots & u^d_{T-L+1} \\
                    u_2^d & u^d_3 & \dots & u^d_{T-L+2} \\
                    \vdots & \vdots &\ddots &\vdots \\
                    u^d_{L} & u^d_{L+1} & \dots & u^d_{T}
                    \end{array}\right]
\end{equation}
The sequence $\mathbf{u}_T^d$ is said to be persistently exciting of order $L$, if $\mathscr{H}_L(\mathbf{u}_T^d)$ has full row rank.
\end{definition}

With a persistently exciting input sequence $\mathbf{u}_{T}^d$, any input and output trajectories of system~(\ref{ed:linearmodel}) with length $L$, which are denoted by $\mathbf{u}_{L}:=\{u\}_{1}^{L}$ and $\mathbf{y}_{L}:=\{y\}_1^{L}$, can be described by the Hankel matrices $\mathscr{H}_L(\mathbf{u}_{T}^d)$ and $\mathscr{H}_L(\mathbf{y}_{T}^d)$ and a column vector $g$, as detailed in Lemma~\ref{e:Willems}.

\begin{lemma}[Willems' fundamental lemma\cite{1428856,van2020willems}]\label{e:Willems}
Consider a controllable LTI system (\ref{ed:linearmodel}), and consider the case when $\mathbf{u}_{T}^d$ is persistently exciting of order $L+n_x$. Any $L$-step trajectories $\mathbf{u}_{L}:=\{u\}_{1}^{L}\in \mathbb R^{n_uL}$ and $\mathbf{y}_{L}:=\{y\}_1^{L}\in \mathbb R^{n_yL}$ are the input and output trajectories of the LTI system in (\ref{ed:linearmodel}), if and only if
\begin{equation}\label{fl}
    \left[\begin{array}{c}
           \mathscr{H}_L(\mathbf{u}_{T}^d) \\
           \mathscr{H}_L(\mathbf{y}_{T}^d)
          \end{array}
    \right] g = 
    \left[\begin{array}{c}
         \mathbf{u}_{L} \\
         \mathbf{y}_{L} 
    \end{array}
    \right]
\end{equation}
holds for column vector $g\in\mathbb{R}^{T-L+1}$.
\label{deepc:lem}\leavevmode\unskip
\end{lemma}

Based on the Willems' fundamental lemma~\cite{1428856}, data-enabled predictive control (DeePC), which is a data-based input-output control method, was proposed in~\cite{coulson2019data}. Let $T_{ini}, N_p \in \mathbb{N}^+$ and $L=T_{ini}+N_p$. The two Hankel matrices can be divided into the past and the future segments as follows:
\begin{equation}\label{deepc:4}
    \left[\begin{array}{c}
            U_p \\
            U_f
          \end{array}
    \right] := \mathscr{H}_L(\mathbf{u}^d_T),\ \left[\begin{array}{c}
            Y_p \\
            Y_f
          \end{array}
    \right] := \mathscr{H}_L(\mathbf{y}^d_T)
\end{equation}
where the past data $U_p$ and $Y_p$ consist of the first $n_u \times T_{ini}$ rows of $\mathscr{H}_L(\mathbf{u}^d_T)$ and the first $n_y \times T_{ini}$ rows of $\mathscr{H}_L(\mathbf{y}^d_T)$, respectively; the future data $U_f$ and $Y_f$ correspond to the last $n_u\times N_p$ rows of $\mathscr{H}_L(\mathbf{u}^d_T)$ and the last $ n_y\times N_p$ rows of $\mathscr{H}_L(\mathbf{y}^d_T)$, respectively.

According to~\cite{coulson2019data}, with $T_{ini}$-step input and output sequences $\mathbf{u}_{ini, k}:=\{u\}_{k-T_{ini}}^{k-1}$ and $\mathbf{y}_{ini, k}:=\{y\}_{k-T_{ini}}^{k-1}$, $N_p$-step future input and output sequences $\hat{\mathbf{u}}_k:=\{\hat{u}\}_{k|k}^{k+N_p-1|k}$ and $\hat{\mathbf{y}}_k:=  \{\hat{y}\}_{k|k}^{k+N_p-1|k}$ can be predicted. The corresponding optimization problem at the time instant $k$ can be formulated as follows~\cite{coulson2019data}:
\begin{subequations}\label{deepc:deepc_opt}
\begin{align}
        \min_{g_k, \hat{\mathbf{u}}_k, \hat{\mathbf{y}}_k} \  &\Vert \hat{\mathbf{y}}_k - \mathbf{y}^r_k \Vert_T^2 + \Vert \hat{\mathbf{u}}_k -\mathbf{u}^r_k\Vert_R^2 \label{deepc:deepc_opt:prob}\\
        \text{s.t.} \quad &\left[\begin{array}{c}
            U_p \\
            Y_P \\
            U_f \\
            Y_f
          \end{array}
    \right] g_k = 
    \left[\begin{array}{c}
         \mathbf{u}_{ini,k} \\
         \mathbf{y}_{ini,k} \\
         \hat{\mathbf{u}}_k \\
         \hat{\mathbf{y}}_k
    \end{array} 
    \right]\label{deepc:deepc_opt:1} \\
    &\hat{u}_{j|k} \in \mathbb{U}, \quad j = k, \ldots, k+N_p-1\\
    &\hat{y}_{j|k} \in \mathbb{Y}, \quad j = k, \ldots, k+N_p-1 
\end{align}
\end{subequations}
where $\mathbf{u}^r_k:=\{u^r\}_k^{k+N_p-1}$ and $\mathbf{y}^r_k:=\{y^r\}_k^{k+N_p-1}$ are the input and output references; $T\in \mathbb{R}^{n_yN_p\times n_yN_p}$ and $R\in \mathbb{R}^{n_uN_p\times n_uN_p}$ are the weighting matrices. The first element $\hat u^*_{k|k}$ of the optimal input sequence $\hat{\mathbf{u}}_k^* = \big[\hat{u}^{* \top}_{k|k}, \ldots, \hat{u}_{k+N_p-1|k}^{* \top}\big]^\top$ is applied to the LTI system~(\ref{ed:linearmodel}) at the time instant $k$.

\subsection{Koopman operator}
Consider a general discrete-time autonomous nonlinear system as follows:
\begin{equation}
\label{ed:K:nonliear}
    x_{k+1} = f(x_k)
\end{equation}
where $k$ denotes the sampling instant; $x \in \mathbb{X} \subseteq \mathbb{R}^{n_x} $ is the system state.

Based on the Koopman operator theory~\cite{koopman1931hamiltonian, budivsic2012applied}, a linear representation of system~(\ref{ed:K:nonliear}) can be determined within a lifted space $\mathcal H(\mathbb X)$, which is a Hilbert space containing all square-integrable real-valued functions defined over $\mathbb X$. The infinite-dimensional lifted state $\Psi(x_k)$ evolves linearly within the lifted space $\mathcal H(\mathbb X)$, governed by the Koopman operator $\mathcal K: \mathcal H(\mathbb X)\to \mathcal H(\mathbb X)$ as follows:
\begin{equation}
    \Psi( f(x_k) )= \mathcal K \Psi(x_k)
\end{equation}
where the lifting function $\mathbf \Psi(x)$ is typically complete or at least chosen to include all components of interest~\cite{brunton2021modern}. In practical applications, the lifting function $\Psi(\cdot)$ and the Koopman operator $\mathcal K$ are often approximated by finite-dimensional representations~\cite{korda2018linear, han2020deep, lusch2018deep}.

The Koopman operator has been extended to address general nonlinear controlled systems with the following form~\cite{korda2018linear}:
\begin{equation}
\label{ed:K:nonliear_2}
    x_{k+1} = f(x_k,u_k)
\end{equation}
where $u \in \mathbb{U} \subseteq \mathbb{R}^{n_u}$ represents the control input; $f: \mathbb{X}\times\mathbb{U}\to\mathbb{X}$ characterizes the nonlinear dynamics. An augmented state vector $\mathcal X_k = [x_k^\top\ \mathbf u_k^\top]^\top$ is constructed, where $ \mathbf u_k = \{u\}_{k}^\infty$ denotes the sequence of control inputs from time instant $k$ onward~\cite{korda2018linear}. $\hat \Psi(\mathcal X_k)= [\Phi(x_k)^\top\ u_k^\top]^\top$, where $\Phi(x_k)\subseteq \mathbb R^{n_s}$ is the lifted state, is used as the augmented state vector. This formulation disregards future control inputs and input-state interactions, yet it facilitates the design of a finite-dimensional Koopman operator $\hat {\mathcal K}$. Only the first $n_s$ rows of $\hat {\mathcal K}$ are needed for predicting the time evolution of $\Phi(x)$. Denoting the first $n_s$ rows of $\hat {\mathcal K}$ as $[ A_{\mathcal K},  B_{\mathcal K}]$, a finite-dimensional approximate Koopman model takes the following form~\cite{korda2018linear}:
\begin{equation}\label{ed:k:model1}
    \Phi(x_{k+1}) = A_{\mathcal K} \Phi(x_k) +B_{\mathcal K}u_k
\end{equation}
where $A_{\mathcal K}\in \mathbb R^{n_s\times n_s}$ and $B_{\mathcal K}\in \mathbb R^{n_s\times n_u}$ are sub-matrices of the approximated Koopman operator $\hat{\mathcal K}$~\cite{korda2018linear}.

\subsection{Problem statement}\label{sec:prob}
In this work, we consider a class of general discrete-time nonlinear systems as follows:
\begin{subequations}\label{ed:nlmodel}
    \begin{align}
        x_{k+1} &= f(x_k, u_k)\label{ed:nlmodel:1}\\
        y_k &= Cx_k\label{ed:nlmodel:2}
    \end{align}
\end{subequations}
where $x \in \mathbb{X} \subseteq \mathbb{R}^{n_x} $ is the system state vector; $y \in \mathbb{Y} \subseteq \mathbb{R}^{n_y}$ is the vector of measured outputs; $u \in \mathbb{U} \subseteq \mathbb{R}^{n_u} $ is the system input vector; 
$C \subseteq \mathbb{R}^{n_y \times n_x}$ is the output measurement matrix.

For nonlinear system~(\ref{ed:nlmodel}), we consider a general economic cost function $\ell_e: \mathbb{Y}\times\mathbb{U}\to\mathbb{C}$, which describes the real-time economic operational cost as a function of the current system outputs and inputs, is with the following general form:
\begin{equation}\label{ed:ecost}
    c_k = \ell_e(y_k, u_k) = \ell_y(y_k)+\ell_u(u_k)
\end{equation}
where $c_k \in \mathbb{C} \subseteq \mathbb{R} $ denotes the economic cost at time instant $k$.

In this work, we aim to propose a data-enabled economic predictive control (DeeEPC) approach by leveraging the DeePC framework introduced in~\cite{coulson2019data}. The control objective is to minimize the economic cost in~(\ref{ed:ecost}) while satisfying hard constraints on the system outputs and control inputs. We consider the practical scenario where no accurate dynamic model is available to describe the nonlinear system in~(\ref{ed:nlmodel}), and only historical input-output data are accessible. In this context, instead of designing a nonlinear EMPC controller following representative model-based approaches (e.g.,~\cite{rawlings2012fundamentals, faulwasser2018economic, ellis2014tutorial, zeng2015economic}), we aim to leverage the DeePC framework~\cite{coulson2019data} to develop a data-driven economic predictive control method. Additionally, despite the nonlinearity in the cost function in~(\ref{ed:ecost}), our goal is to preserve the convexity of the formulated optimization problem, which is a favorable characteristic of DeePC~\cite{coulson2019data}.

\begin{figure*}[t!]
    \centering
    \includegraphics[width=0.8\textwidth]{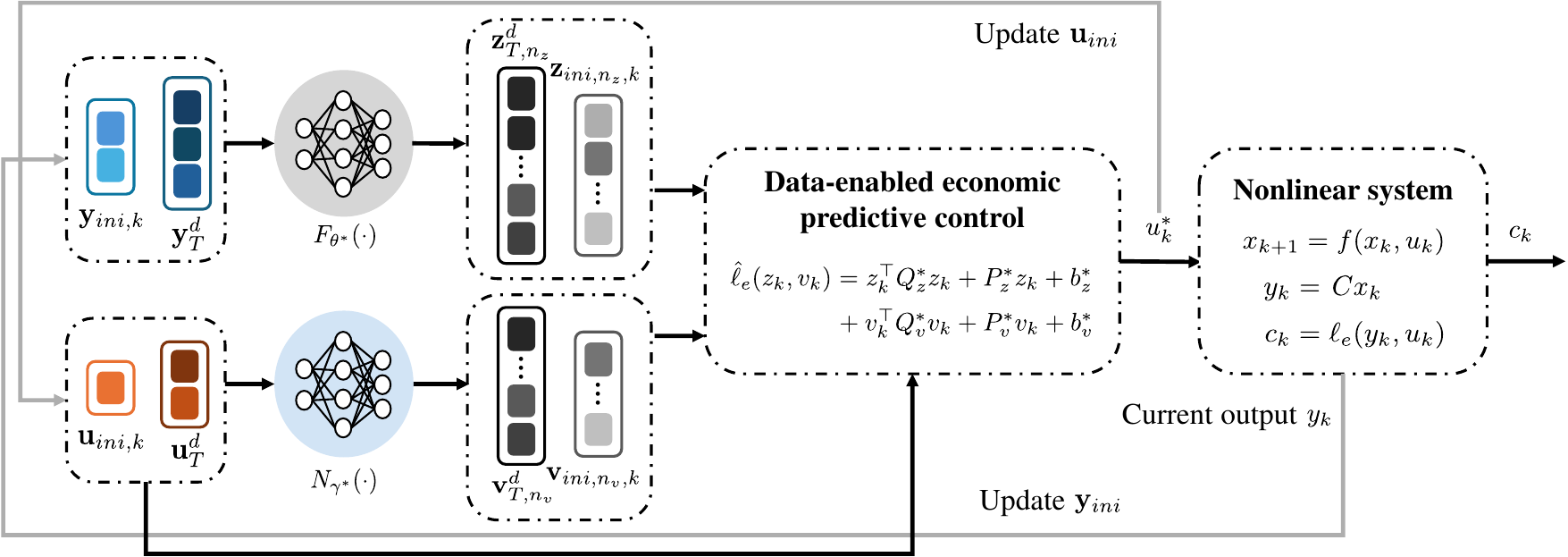}
    \caption{Architecture of the proposed data-enabled economic predictive control based on convex optimization.}
    \label{ed:fig:control}
\end{figure*}

\begin{remark}
The economic cost function in~(\ref{ed:ecost}) represents a general class of economic performance indicators commonly encountered in various applications (e.g., \cite{heidarinejad2012economic,wu2020economic,lao2014economic,du2017real,haque2020advanced,  nonhoff2019economic,huang2022economic}), including the two case studies~\cite{zeng2015economic,zhang2025machine} considered in this paper.
\end{remark}

\section{Architecture of the proposed DeeEPC}\label{sec:mian_design}

While the objective is to minimize the economic cost, directly replacing the quadratic cost function in~(\ref{deepc:deepc_opt:prob}) with the general economic cost function in~(\ref{ed:ecost}) will compromise the convexity of the original DeePC formulation in~\cite{coulson2019data}, when $\ell_e$ in~(\ref{ed:ecost}) is nonlinear. To address this, the proposed DeeEPC employs two nonlinear mappings, $\Phi_z$ and $\Phi_v$, which compute the mapped output $z_k$ and mapped input $v_k$ respectively, expressed as $z_k= \Phi_z (y_k)$ and $v_k= \Phi_v (u_k)$. 
The finite-dimensional approximations of these two nonlinear mappings are realized using neural networks (denoted as $F_\theta(\cdot)$ and $N_\gamma(\cdot)$ in Figure~\ref{ed:fig:control}).
With the mapped output $z_k$ and the mapped input $v_k$, based on the economic cost approximation designs in~\cite{han2024efficient,han2025deep}, we construct a quadratic approximation of the nonlinear economic cost function~(\ref{ed:ecost}). This approximation, denoted by $\hat \ell_e(z_k,v_k)$, has the following form:
\begin{equation}\label{ed:cost_qua}
    \hat \ell_e(z_k,v_k) = \underbrace{
    z_k^\top Q_z z_k + P_z z_k + b_z
}_{\hat \ell_y(z_k)} + \underbrace{v_k^\top Q_v v_k + P_vv_k + b_v}_{\hat \ell_u(v_k)}
\end{equation}
where $Q_z\in\mathbb{R}^{n_z\times n_z}$ and $Q_v\in\mathbb{R}^{n_v\times n_v}$ are positive definite matrices; $P_z\in\mathbb{R}^{1\times n_z}$ and $P_v\in\mathbb{R}^{1\times n_v}$ are real-valued vectors; $b_z\in\mathbb{R}$ and $b_v\in\mathbb{R}$ are scalars. 

Based on the nonlinear mappings $\Phi_z$ and $\Phi_v$, and the quadratic economic cost function $\hat \ell_e$ in~(\ref{ed:cost_qua}), the data-enabled economic predictive control (DeeEPC) design can be formulated as the following convex optimization problem:
\begin{subequations}\label{deepc:keconomic DeePC_opt}
\begin{align}
    \min_{g_k} \sum_{j=k}^{k+N_p-1}\ & \lambda(\hat z_{j|k}^\top Q_z\hat z_{j|k} + P_z\hat z_{j|k}+b_z \notag \\[-2.2ex]
    &  + \hat v_{j|k}^\top Q_v\hat v_{j|k} + P_v\hat v_{j|k}+b_v) \label{deepc:keconomic DeePC_opt:prob} \allowdisplaybreaks\\ 
    \text{s.t.}\ &\left[\begin{array}{c}
            U_p \\
            V_p \\
            Z_p \\
            U_f \\
            V_f \\
            Z_f
          \end{array}
    \right]\ g_k = 
    \left[\begin{array}{c}
         \mathbf u_{ini,k} \\
         \mathbf v_{ini,k} \\
         \mathbf z_{ini,k} \\
         \hat{\mathbf{u}}_k \\
         \hat{\mathbf{v}}_k \\
         \hat{\mathbf{z}}_k
    \end{array}
    \right]\label{deepc:keconomic DeePC_opt:1}\\
    &\hat{u}_{j|k}\in\mathbb{U},\ j = k,\dots,k+N_p-1\label{deepc:keconomic DeePC_opt:2} \\
    &G\hat{z}_{j|k} \in \mathbb{Y}_c,\ j = k,\dots,k+N_p-1\label{deepc:keconomic DeePC_opt:3}
\end{align}
\end{subequations}
In~(\ref{deepc:keconomic DeePC_opt:prob}), $\lambda \in \mathbb{R}$ is a positive scalar.
In~(\ref{deepc:keconomic DeePC_opt:1}), $\mathbf{z}_{ini, k}:=\{z\}_{k-T_{ini}}^{k-1}$ and $\mathbf{v}_{ini, k}:=\{v\}_{k-T_{ini}}^{k-1}$ are the initialization trajectories of the mapped output and mapped input, respectively; $\hat{\mathbf{z}}_k:=  \{\hat{z}\}_{k|k}^{k+N_p-1|k}$ and $\hat{\mathbf{v}}_k:=  \{\hat{v}\}_{k|k}^{k+N_p-1|k}$ are the predicted future mapped output and mapped input trajectories, respectively. 
$Z_p$ and $Z_f$ are the segments of the Hankel matrix constructed using the offline-collected mapped output, that is, $[Z_p^\top , Z_f^\top ]^\top:= \mathscr H_{Tini+Np}(\mathbf z_T^d)$ where $\mathbf z_T^d:=\{z^d\}_{1}^{T}$. Similarly, $V_p$ and $V_f$ are the segments of the Hankel matrix $\mathscr H_{Tini+Np}(\mathbf v_T^d)$ constructed using the offline-collected mapped input.
(\ref{deepc:keconomic DeePC_opt:2}) handles input constraints. (\ref{deepc:keconomic DeePC_opt:3}) handles hard constraints on $y^c$, which is a subset of the measured output vector $y$; specifically, $\mathbb Y_c$ is a compact set that $y^c$ must belong to, i.e., $y^c\in \mathbb Y_c$. $G$ is an output reconstruction matrix that predicts $y_c$ from the mapped outputs.
The original input is included in (\ref{deepc:keconomic DeePC_opt:1}) to preserve the convexity of the formulated optimization problem and to handle hard input constraints. Specifically, the input applied to the system at each sampling instant is obtained as the first element of the optimal input sequence $\hat{\mathbf{u}}_k^* = U_fg_k^*$. Input constraints are enforced by (\ref{deepc:keconomic DeePC_opt:2}), where the input predictions $\hat{u}_{j|k}$ are computed in the fourth line of (\ref{deepc:keconomic DeePC_opt:1}).

Figure~\ref{ed:fig:control} illustrates the proposed data-enabled economic predictive control method for nonlinear systems. Finite-dimensional mapped output $z_{n_z}\subseteq\mathbb R^{n_z}$ and mapped input $v_{n_v}\subseteq\mathbb R^{n_v}$ are obtained using the trained neural networks $F_\theta(\cdot)$ and $N_\gamma(\cdot)$, and are subsequently used to formulate the optimization problem~(\ref{deepc:keconomic DeePC_opt}). The parameters in~(\ref{ed:cost_qua}) and the reconstruction matrix in~(\ref{deepc:keconomic DeePC_opt:3}) are learned from data.



\begin{remark}
    The structure of $\hat \ell_e$ in~(\ref{ed:cost_qua}), used to approximate nonlinear economic cost function $\ell_e$ in~(\ref{ed:ecost}), is inspired by the designs presented in~\cite{han2024efficient,han2025deep} (specifically, Section 4.2 of~\cite{han2024efficient} and Section III-C of~\cite{han2025deep}), and has been adopted in the conference version of this work~\cite{yan2025economic}.
    Meanwhile, in contrast to the design in~\cite{yan2025economic}, where the approximated economic cost depends only on the mapped outputs, the approximated economic cost function in~(\ref{ed:cost_qua}) is constructed to depend on both the mapped outputs and the mapped inputs. This extension allows the proposed method to effectively handle a broader range of nonlinear systems with economic costs that are influenced by both outputs and inputs, including the water treatment and carbon capture processes discussed in Section~\ref{sec:wwtp}.
\end{remark}

\section{Rationale of using input and output mappings}
\label{sec:proof}
In this section, we justify the use of nonlinear mappings in the proposed DeeEPC as a means to effectively manage nonlinearities in the system and its economic cost function, based on the Koopman-based modeling framework~\cite{korda2018linear}.

\subsection{Expected properties of Koopman-compatible nonlinear mappings for DeeEPC}\label{sec:req}


To align the optimization problem~(\ref{deepc:keconomic DeePC_opt}) with Willems’ fundamental lemma~\cite{1428856}, we require that the mapped output vector $z$ and the mapped input vector $v$ belong to a virtual LTI system.

According to the Koopman modeling approach in~\cite{korda2018linear}, the nonlinear system in~(\ref{ed:nlmodel:1}) can be approximately represented by a linear Koopman model in the form of~(\ref{ed:k:model1}). 
The modeling framework in~\cite{korda2018linear} has been extended in~\cite{shi2022deep,li2024machine}, where nonlinear mapping of the original state and joint mapping of both the original state and known input are considered. Partially inspired by~\cite{shi2022deep,li2024machine}, a modified Koopman model of system~(\ref{ed:nlmodel:1}) may be expressed as follows:
\begin{equation}\label{ed:model_A}
\begin{aligned}
    \Phi(x_{k+1}) &= A_{\mathcal K} \Phi(x_k) + B_1 u_k + B_2 \Phi_v(u_k) \\
                  &= A_{\mathcal K} \Phi(x_k) + \underbrace{[B_1\ B_2]}_{B_{\mathcal K}}  
                     \begin{bmatrix}
                         u_k \\
                         v_k
                     \end{bmatrix}
\end{aligned}
\end{equation}
where $\Phi_v(\cdot)$ is the input lifting function, $v_k= \Phi_v(u_k)$ is the mapped input that characterizes the nonlinear dependence of the lifted state on the control input. 

With a model structure similar to the Koopman-based structures in~\cite{shi2022deep,li2024machine} (except that another nonlinear mapping that captures the coupling between state and input is explicitly considered in~\cite{shi2022deep,li2024machine}), the modified Koopman model structure~(\ref{ed:model_A}) is expected to be a good approximation of the nonlinear system in~(\ref{ed:nlmodel}).
The model accuracy of~(\ref{ed:model_A}) is expected to be no worse than~(\ref{ed:k:model1}) proposed in~\cite{korda2018linear}, since the modified Koopman model in the form of~(\ref{ed:model_A}) further incorporates the nonlinear mapping of the control input $u$. We now introduce the definition of complete orthonormal system.

\begin{definition}[Complete orthonormal system~\cite{kashin2005orthogonal}]\label{ed:def:co}
    Consider a one-dimensional variable $x\in X\subseteq \mathbb{R}$. 
    Let $\mu$ denote the Lebesgue measure on $X$, the space $L^2(X,\mu)$ is a Hilbert space containing the measurable function $h(x)\in L^2(X,\mu)$ satisfying:
    \begin{equation}
        \Vert h(x) \Vert = \sqrt{\int_X |h(x)|^2dx }<\infty 
    \end{equation}
    where $dx:=d\mu(x)$. 

    
     A sequence of basis functions $\{\phi_i(x)\}_{i=1}^\infty$ in $L^2(X,\mu)$ forms a complete orthonormal system (C.O.N.S.), if and only if it satisfies the following properties:
    \begin{itemize}
        \item Completeness: For any $h(x)\in L^2(X,\mu)$, $ h(x) = \sum\limits_{i=1}^\infty c_i  \phi_i(x) $, where $c_i = \langle h(x), \phi_i(x)\rangle$.
        \item Orthogonality: $\langle \phi_i(x), \phi_k(x) \rangle = \int_{X}\phi_i(x)\phi_k(x)dx = \delta_{ik}$, where \( \delta_{ik} \) is the Kronecker delta.
        \item Normalization: $\Vert \phi_i(x) \Vert^2 = \int_X |\phi_i(x)|^2dx =1 $.
    \end{itemize}
\end{definition}
Lebesgue measure is adopted throughout the remainder of the paper, and for any $X_i\in\mathbb{R}$, $L^2(X_i, \mu_i)$ will be abbreviated as $L^2(X_i)$ for brevity.

A common choice of Koopman modeling is to construct the lifting function as a complete orthonormal system (C.O.N.S.) in an $L^2$ space to ensure lifted states can effectively approximate general functions~\cite{brunton2021modern, linares2019koopman, servadio2023koopman}. In this work, we consider that the nonlinear mappings $\mathbf \Phi(x)=\{\phi_i(x)\}_{i=1}^\infty$ and $\mathbf \Phi_v(u)=\{\phi_{j}^v(u)\}_{j=1}^\infty$, which are used to construct the lifted state $\Phi(x_k)$ and the mapped input $\Phi_v(u_k)$ in~(\ref{ed:model_A}), are C.O.N.S. in $L^2(\mathbb X)$ and $L^2(\mathbb U)$, respectively.

The Koopman model structure in~(\ref{ed:model_A}) will be leveraged as the basis for illustrating the rationale behind the proposed DeeEPC framework presented in Section~\ref{sec:mian_design}. Specifically, we aim to prove the existence of nonlinear mappings
$\mathbf \Phi_z(y)$ and $\mathbf \Phi_v(u)$ that satisfy the following two conditions simultaneously.

\begin{itemize}
    \item Condition 1: The mapped output $z$ and the augmented input $[u^\top, v^\top]^\top$ belong to a virtual Koopman-based model in the following form:
    \begin{subequations}\label{ed:model_B}
    \begin{align}
        \Phi(x_{k+1}) &= A_{\mathcal K} \Phi(x_{k}) +B_{\mathcal K}[u_k^\top, v_k^\top]^\top\label{ed:model_B:1}\\
        z_k & = \Phi_z(y_k) =E\Phi(x_{k})\label{ed:model_B:2}
    \end{align}
    \end{subequations}
    where $\Phi_z(\cdot)$ is the output lifting function which generates the mapped output $z_k$, whose elements correspond to the functions in the output mapping $\mathbf \Phi_z(\cdot)$. $E$ characterizes the linear dependence of $z_k$ on the lifted state.
    \item Condition 2: The economic cost function $\ell_e(y,u)$ in~(\ref{ed:ecost}) can be equivalently expressed as $\hat{\ell}_e(z,v)$ in~(\ref{ed:cost_qua}), that is, $\ell_e(y,u) =\hat{\ell}_e(z,v)$, where $z$ and $v$ in~(\ref{ed:cost_qua}) are obtained based on the output and input mappings $\mathbf \Phi_z(y)$ and $\mathbf \Phi_v(u)$, respectively. 
\end{itemize}


\begin{assumption}\label{ed:ass:3}
For the $i$th element of the system state vector $x = [x_{(1)},\dots,x_{(n_x)}]^\top$, the domain of which is denoted as $X_{i}$, there exists a C.O.N.S. $\mathbf{\Phi}_i(x_{(i)}) = \{\phi_{i,j}(x_{(i)})\}_{j=1}^\infty$ in $L^2(X_{i})$, $i=1,\dots,n_x$.
There exists a C.O.N.S. in  $L^2(\mathbb U)$ in the form of $\mathbf{\Phi}_{v}(u)=\{\phi_{j}^v(u)\}_{j=1}^\infty$.
\end{assumption}

\begin{assumption}\label{ed:ass:2}
    Let $n_r$ denote the number of non-zero columns of the output measurement matrix $C$. When $n_y<n_x$, $C$ satisfies the following properties:
    \begin{itemize}
        \item The number of non-zero columns of $C$ is no greater than the number of rows, i.e., $n_r\leq n_y$;
        \item The non-zero columns of $C$ are linearly independent, i.e., $\mathrm{rank}(C) =n_r$.
    \end{itemize}
    When $n_y\geq n_x$, this assumption reduces to $\mathrm{rank}(C) = n_x$.
\end{assumption}

\begin{assumption}\label{ed:ass:1}
    Both $\ell_y(y)$ and $\ell_u(u)$ in~(\ref{ed:ecost}) are square-integrable real-valued functions, that is, $\ell_y(y)\in L^2(\mathbb{Y})$ and $\ell_u(u)\in L^2(\mathbb{U})$.
\end{assumption}

The proposed DeeEPC framework is applicable to economic control of a broad class of nonlinear systems that can be described in the form of~(\ref{ed:nlmodel}) and satisfy two requirements: 1) the measurement matrix $C$ in (\ref{ed:nlmodel:2}) satisfies Assumption~\ref{ed:ass:2}; 2) the economic cost function is assumed not to contain coupling terms between the measurable output $y$ and the control input $u$, as described in~(\ref{ed:ecost}).

\subsection{Existence of $\mathbf \Phi_z(y)$ and $\mathbf \Phi_v(u)$}
\label{sec:existence}
 

In this section, we provide a specific choice of $\mathbf \Phi_z(y)$ and $\mathbf \Phi_v(u)$ to demonstrate their existence. Specifically, a state mapping $\mathbf \Phi(x)$ that forms a C.O.N.S. in $L^2(\mathbb X)$ is designed and serves to define the lifted state $\Phi(x)$. $\mathbf \Phi_z(y)$ is selected based on a subsequence of $\mathbf \Phi(x)$. $\mathbf{\Phi}_{v}(u)$ is chosen as a C.O.N.S. in $L^2(\mathbb U)$, as defined in Assumption~\ref{ed:ass:3}. In the following, we introduce some propositions, which facilitate the selection of $\mathbf{\Phi}_{z}(y)$.


\begin{Proposition}\label{ed:thm:1}
If Assumption~\ref{ed:ass:3} holds,
consider the system state vector $x = [x_{(1)},\dots,x_{(n_x)}]^\top \in \mathbb{X} \subseteq  \mathbb{R}^{n_x}$, the product of $\mathbf{\Phi}_i(x_{(i)}) = \{\phi_{i,j}(x_{(i)})\}_{j=1}^\infty$, $i=1,\dots,n_x$, denoted as $\mathbf{\Phi}(x) = \bigotimes_{i=1}^{n_x} \mathbf{\Phi}_i(x_{(i)})$, is a C.O.N.S. in $L^2(\mathbb{X})$.
\end{Proposition}
\textbf{Proof.} 
The proof of this proposition is inspired by Section II.4 in~\cite{reed1980}. If Assumption~\ref{ed:ass:3} holds, there exists $\mathbf{\Phi}_i(x_{(i)})$, which constitutes a C.O.N.S. in $L^2(X_i)$, $i=1,\dots,n_x$. 
We begin by considering the case $n_x=2$ with the state vector $x = [x_{(1)},x_{(2)}]^\top \in X_1\times X_2$. Under this setting, we show that $\mathbf{\Phi}_1(x_{(1)})\otimes \mathbf{\Phi}_2(x_{(2)})$ is a C.O.N.S.  in $L^2( X_1\times X_2)$. According to Definition~\ref{ed:def:co}, the following equations hold:
\begin{subequations}\label{ed:cons:ortho}
    \begin{align}
        \langle \phi_{1,i}, \phi_{1,k} \rangle &= \int_{X_1}\phi_{1,i}(x_{(1)})\phi_{1,k}(x_{(1)})dx_{(1)} = \delta_{ik}\\
        \langle \phi_{2,j}, \phi_{2,l} \rangle &= \int_{X_2} \phi_{2,j}(x_{(2)})\phi_{2,l}(x_{(2)})dx_{(2)} = \delta_{jl}
    \end{align}
\end{subequations}
and 
\begin{subequations}\label{ortho}
    \begin{align}
        \int_{X_1}|\phi_{1,i}(x_{(1)})|^2dx_{(1)} &= 1\label{ortho:h}\\
        \int_{X_2}|\phi_{2,j}(x_{(2)})|^2dx_{(2)} & = 1\label{ortho:g}
    \end{align}
\end{subequations}
The product 
$\mathbf{\Phi}([x_{(1)},x_{(2)}]^\top)=\{\phi_{ij}(x_{(1)},x_{(2)}): i \in \mathbb N^+,j \in \mathbb N^+\}$ 
consists of elements $\phi_{ij}(x_{(1)},x_{(2)}) = \phi_{1,i}(x_{(1)})\phi_{2,j}(x_{(2)})$. 

First, we show that $\mathbf{\Phi}([x_{(1)},x_{(2)}]^\top)$ is normalized. Since $|\phi_{ij}(x_{(1)},x_{(2)})|^2 \geq 0$, according to the Tonelli's Theorem~\cite{folland1999real}, the square of the $L_2$-norm of $\phi_{ij}(x_{(1)},x_{(2)})$ can be computed as:
\begin{subequations}\label{normalized}
{\small
    \begin{align}
        \Vert \phi_{ij}(x_{(1)},x_{(2)}) \Vert^2 &= \int_{X_1\times X_2}|\phi_{ij}(x_{(1)},x_{(2)})|^2dx\\
        & = \int_{X_1}\int_{X_2}|\phi_{1,i}(x_{(1)})|^2|\phi_{2,j}(x_{(2)})|^2dx_{(2)}dx_{(1)}\\
        & = \int_{X_1}|\phi_{1,i}(x_{(1)})|^2dx_{(1)}\int_{X_2}|\phi_{2,j}(x_{(2)})|^2dx_{(2)} \overset{(\ref{ortho})}{=} 1\label{normalized:3}
    \end{align}   }
\end{subequations}

Next, we prove the elements of $\mathbf{\Phi}([x_{(1)},x_{(2)}]^\top)$ are mutually orthogonal. For any $i,j,k,l\in\mathbb N^+$, two elements of $\mathbf{\Phi}([x_{(1)},x_{(2)}]^\top)$, denoted as $\phi_{ij}(x_{(1)},x_{(2)})$ and $\phi_{kl}(x_{(1)},x_{(2)})$, are defined as follows:
\begin{subequations}
\begin{align}
    \phi_{ij}(x_{(1)},x_{(2)}) &= \phi_{1,i}(x_{(1)})\phi_{2,j}(x_{(2)})\label{el1}\\
    \phi_{kl}(x_{(1)},x_{(2)}) &= \phi_{1,k}(x_{(1)})\phi_{2,l}(x_{(2)})\label{el2}
    \end{align}
\end{subequations}
When $i=k$ and $j=l$, (\ref{el1}) and (\ref{el2}) correspond to the same element of $\mathbf{\Phi}([x_{(1)},x_{(2)}]^\top)$.
Since $\phi_{ij}(x_{(1)},x_{(2)})$ and $\phi_{kl}(x_{(1)},x_{(2)})$ are both square-integrable by~(\ref{normalized}), the following bound can be derived based on the Cauchy-Schwarz inequality~\cite{steele2004cauchy}:
\begin{subequations}
{\small
    \begin{align}
        &\int_{X_1\times X_2}  \vert \phi_{ij}(x_{(1)},x_{(2)})\phi_{kl}(x_{(1)},x_{(2)})\vert dx\\ 
        &= \int_{X_1\times X_2} \vert \phi_{ij}(x_{(1)},x_{(2)})\vert \vert \phi_{kl}(x_{(1)},x_{(2)})\vert dx\\
        &\leq \left(\int_{X_1\times X_2} |\phi_{ij}(x_{(1)},x_{(2)})|^2 dx\right)^{\frac{1}{2}} \left(\int_{X_1\times X_2} |\phi_{kl}(x_{(1)},x_{(2)})|^2 dx\right)^{\frac{1}{2}}\\
        &\overset{(\ref{normalized})}{=} 1
    \end{align} }
\end{subequations}
This shows that $\phi_{ij}(x_{(1)},x_{(2)})\phi_{kl}(x_{(1)},x_{(2)})$ is Lebesgue integrable. Consequently, based on Fubini's theorem~\cite{folland1999real}, the inner product of $\phi_{ij}(x_{{(1)}},x_{(2)})$ and $\phi_{kl}(x_{(1)},x_{(2)})$ can be calculated as follows:
\begin{subequations}\label{ed:ortho}
    \begin{align}
        &\langle \phi_{ij}(x_{(1)},x_{(2)}) , \phi_{kl}(x_{(1)},x_{(2)}) \rangle \\
        &= \int_{X_1}\int_{X_2}\phi_{1,i}(x_{(1)})\phi_{2,j}(x_{(2)})\phi_{1,k}(x_{(1)})\phi_{2,l}(x_{(2)})dx_{(2)}dx_{(1)} \allowdisplaybreaks\\
       &= \int_{X_1}\phi_{1,i}(x_{(1)})\phi_{1,k}(x_{(1)})dx_{(1)}\int_{X_2}\phi_{2,j}(x_{(2)})\phi_{2,l}(x_{(2)})dx_{(2)} \label{ortho:2}\\
       & \overset{(\ref{ed:cons:ortho})}{=} \delta_{ik}\delta_{jl}
    \end{align}
\end{subequations}
When $i=k$ and $j=l$, $\langle \phi_{ij}(x_{(1)},x_{(2)}) , \allowbreak \phi_{kl}(x_{(1)},x_{(2)}) \rangle = 1$ according to~(\ref{ed:ortho}), which implies the constructed system is normalized.

Finally, we show that any function $\xi(x_{(1)},x_{(2)})\in L^2(X_1\times X_2)$ can be represented by a linear combination of elements in $\mathbf{\Phi}([x_{(1)},x_{(2)}]^\top)$. We fix $x_{(2)}=x_{(2)}^*$. Then, for each fixed $ x_{(2)}^*\in X_2$, $\xi(x_{(1)},x_{(2)}^*)$ depends only on $x_{(1)}$. Since $\{\phi_{1,i}(x_{(1)})\}_{i=1}^\infty$ is complete, $\xi(x_{(1)},x_{(2)}^*)$ can be represented as follows:
\begin{equation}\label{ed:prof1:4}
    \xi(x_{(1)},x_{(2)}^*) = \sum\limits_{i=1}^\infty c_i\phi_{1,i}(x_{(1)})
\end{equation} 
where the coefficients $c_i = \langle \xi(x_{(1)},x_{(2)}^*),\phi_{1,i}(x_{(1)})\rangle =   \int_{X_1} \xi(x_{(1)}, \\ x_{(2)}^*)\phi_{1,i}(x_{(1)})dx_{(1)}$ depend solely on the fixed value $x_{(2)}^*$. 
The upper bound of the squared integration of $c_i(x_{(2)})$ can be obtained based on the Cauchy-Schwarz inequality~\cite{steele2004cauchy} as follows:
\begingroup
\allowdisplaybreaks
\begin{subequations}\label{ed:ci_L2}
\begin{align}
    &\int_{X_2}|c_i(x_{(2)})|^2dx_{(2)} = \int_{X_2}\big |\int_{X_1}\xi(x_{(1)},x_{(2)})\phi_{1,i}(x_{(1)})dx_{(1)}\big|^2dx_{(2)}\\
    &\leq \int_{X_2}(\int_{X_1}|\xi(x_{(1)},x_{(2)})|^2dx_{(1)}\int_{X_1}|\phi_{1,i}(x_{(1)})|^2dx_{(1)})dx_{(2)}\label{ed:bound_ci}\\
    & = \int_{X_1}|\phi_{1,i}(x_{(1)})|^2dx_{(1)} \int_{X_2}\int_{X_1}|\xi(x_{(1)},x_{(2)})|^2dx_{(1)}dx_{(2)} \\
    & \overset{(\ref{ortho:h})}{=} \int_{X_2}\int_{X_1}|\xi(x_{(1)},x_{(2)})|^2dx_{(1)}dx_{(2)}\label{ed:ci_L2:4}
\end{align}
\end{subequations}
\endgroup
According to Definition~\ref{ed:def:co} and the Tonelli's Theorem~\cite{bauer2001measure}, the right-hand side of~(\ref{ed:ci_L2:4}) is equivalent to $\Vert \xi(x_{(1)},x_{(2)}) \Vert^2$. Since $\xi(x_{(1)},x_{(2)})\in L^2(X_1\times X_2)$, (\ref{ed:ci_L2}) indicates that $\Vert c_i(x_{(2)})\Vert^2<\infty$, i.e., $c_i(x_{(2)})\in L^2(X_2)$ for all $x_{(2)}\in X_2$ according to Definition~\ref{ed:def:co}. Then due to the completeness of $\{\phi_{2,j}(x_{(2)})\}_{j=1}^\infty$, (\ref{ed:prof1:4}) can be rewritten as follows:
\begin{subequations}\label{ed:complete}
    \begin{align}
        \xi(x_{(1)},x_{(2)}) &= \sum\limits_{i=1}^\infty c_i(x_{(2)}) \phi_{1,i}(x_{(1)}) \\
        & = \sum\limits_{i=1}^\infty \sum\limits_{j=1}^\infty b_{ij} \phi_{2,j}(x_{(2)}) \phi_{1,i}(x_{(1)}) \\
        & = \sum\limits_{i=1}^\infty \sum\limits_{j=1}^\infty b_{ij} \phi_{ij}(x_{(1)},x_{(2)})
    \end{align}
\end{subequations}
where $b_{ij} = \langle \xi(x_{(1)},x_{(2)}), \phi_{ij}(x_{(1)},x_{(2)})   \rangle$. (\ref{ed:complete}) indicates $\mathbf{\Phi}([\allowbreak x_{(1)},\allowbreak x_{(2)}]^\top)$ is complete in $L^2(X_1\times X_2)$.


When $n_x =  3$, based on the above analysis, $\mathbf{\Phi}([x_{(1)},x_{(2)},\allowbreak  x_{(3)}]^\top) = \mathbf{\Phi}([x_{(1)},x_{(2)}]^\top)\bigotimes \mathbf{\Phi}_3(x_{(3)})$ is a C.O.N.S. in $L^2(X_1\times X_2 \times X_3)$, since both $\mathbf{\Phi}([x_{(1)},x_{(2)}]^\top)$ and $\mathbf{\Phi}_3(x_{(3)})$ are C.O.N.S.. By mathematical induction, $\mathbf{\Phi}(x) = \bigotimes_{i=1}^{n_x} \mathbf{\Phi}_i(x_{(i)})$ is a C.O.N.S. in $L^2(\mathbb{X})$.
$\square$

Following Proposition~\ref{ed:thm:1}, we choose the state mapping $\mathbf{\Phi}(x)$ in the following form:
\begin{equation}\label{ed:lifting function}
    \mathbf{\Phi}(x) = \bigotimes_{i=1}^{n_x} \mathbf{\Phi}_{i}(x_{(i)})
\end{equation}
Then, the corresponding lifting function ${\Phi}(x)$ in (\ref{ed:model_B}) can be constructed from $\mathbf{\Phi}(x)$.
For each state variable $x_{(i)}\in X_i$, the constant function $1$ is typically included in $\mathbf{\Phi}_{i}(x_{(i)})$, as constant functions naturally belong to $L^2(X_i)$ and often appear as components of more general functions (i.e., functions with constant terms). Therefore, without loss of generality, we define the first element in $\mathbf{\Phi}_{i}(x_{(i)})$ to be the constant function $1$, i.e., $\phi_{i,1}(x_{(i)}) =1$, for $i=1,\dots,n_x$. 
A similar setting is adopted in~\cite{pan2022stochastic}. With this design, we can obtain the following results.

\begin{Proposition}\label{ed:prop:2}
If Assumption~\ref{ed:ass:2} holds, then given the system output $y$ and the measurement matrix $C$, $\operatorname{rank}(C)$ elements of the system state $x$ can be uniquely determined.
\end{Proposition}
\textbf{Proof.} First, consider the case when $n_y< n_x$. Let $C'$ denote a submatrix of $C$ consisting of all nonzero columns of $C$. We partition the state vector $x$ into two subvectors: $x^r$ and $x^n$, where
$x^r\in \mathbb X^r$ contains the elements of $x$ corresponding to the nonzero columns of $C$, and $x^n\in \mathbb X^n$ contains the remaining elements of state $x$. Then, (\ref{ed:nlmodel:2}) can be rewritten as follows:
\begin{equation}\label{ed:decompose_y}
    y = Cx = C'x^r+\mathbf{0}x^n = C'x^r
\end{equation}
If Assumption~\ref{ed:ass:2} holds, the columns of $C'$ are linearly independent, then given $y$ and $C$, vector $x_r$ can be computed as:
\begin{equation} \label{ed:xr}
    x^r = (C'^\top C')^{-1} C'^\top y
\end{equation}



When $n_y > n_x$, we have $C'=C$. Following the same procedure, it holds that $x^r = (C^\top C)^{-1} C^\top y$. When $n_y=n_x$, we have $x^r = C^{-1}y$. $\square$

Let $J=\{1,\dots,n_x\}$ denote the set of indices of all the state variables in $x$. Let $I$ denote the subset of indices of the partial state variables of $x$ that appear in $x_r$.

\begin{Proposition}\label{ed:prop:1}
If Assumptions~\ref{ed:ass:3}-\ref{ed:ass:2} hold, and the C.O.N.S. $\mathbf{\Phi}^r(x^r)$ in $L^2(\mathbb X^r)$, whose elements are $\phi_i^r(x^r)$, $i\in \mathbb N^+$, is constructed as $\mathbf{\Phi}^r(x^r) = \bigotimes_{i\in I} \mathbf{\Phi}_i(x_{(i)})$, then $\mathbf{\Phi}^r(x^r)$ is a subsequence of, or equivalent to, $\mathbf{\Phi}(x)$ in~(\ref{ed:lifting function}).
\end{Proposition}
\textbf{Proof.}
If Assumption~\ref{ed:ass:3} holds, a C.O.N.S. $\mathbf \Phi(x)$ can be constructed as defined in~(\ref{ed:lifting function}). If Assumption~\ref{ed:ass:2} holds, $x^r$ can be determined according to Proposition~\ref{ed:prop:2}.
When $I \subset J$, we can establish a subsequence of $\mathbf{\Phi}(x)$ as follows:
\begin{subequations}\label{ed:proposition 1}
{\small
\begin{align}
     \mathbf{\Phi}'(x) &= \Big\{\big(\bigotimes_{k\in J \setminus I}\{1\}\big) \bigotimes \big( \bigotimes_{i\in I} \{1\}\big)  ,  \big( \bigotimes_{k\in J \setminus I} \{1\} \big) \bigotimes \big(\bigotimes_{i\in I} \{\phi_{i,j}(x_{(i)})\}_{j=2}^\infty \big) \Big\}\label{ed:product:1}\\
     &= \Big\{1 ,  \bigotimes_{i\in I} \{\phi_{i,j}(x_{(i)})\}_{j=2}^\infty \Big\} = \mathbf{\Phi}^r(x^r) 
\end{align}}
\end{subequations}
In~(\ref{ed:proposition 1}), $\mathbf{\Phi}'(x)$ is constructed such that, for variables that appear only in $x$, the constant function $1$ is chosen, while for variables that are shared by both $x$ and $x^r$, arbitrary basis functions are selected as in~(\ref{ed:product:1}). Under this construction, $\mathbf{\Phi}'(x)$ contains the same components as $\mathbf{\Phi}^r(x^r)$ and is a subsequence of $\mathbf{\Phi}(x)$, since $\mathbf{\Phi}(x)$ is obtained by selecting basis functions from both $\mathbf{\phi}_i(x_{(i)})$ for $i \in I$ and $\mathbf{\phi}_k(x_{(k)})$ for $k \in J \setminus I$.
When $I=J$, it follows that $\mathbf{\Phi}^r(x^r) =\mathbf{\Phi}(x)$. $\square$



Based on Propositions \ref{ed:thm:1}-\ref{ed:prop:1}, we define $\phi_{i}^z(y)$, which is the $i$th element of the output mapping $\mathbf{\Phi}_z(y)$, as follows:
\begin{equation}\label{ed:phi_z}
    \phi_{i}^z(y) \triangleq \phi_{i}^r((C'^\top C')^{-1} C'^\top y) \overset{(\ref{ed:xr})}{=} \phi_{i}^r(x^r) , \ i\in\mathbb N^+
\end{equation}

\begin{Proposition}\label{ed:prop:3}
If Assumptions~\ref{ed:ass:3}-\ref{ed:ass:1} hold, $\mathbf{\Phi}_z(y)=\{\phi_{i}^z(y)\}_{i=1}^\infty$ is complete in $L^2(\mathbb Y)$.
\end{Proposition} 
\textbf{Proof.}
If Assumptions~\ref{ed:ass:3}-\ref{ed:ass:1} hold, $\mathbf{\Phi}_z(y)$ can be constructed as defined in (\ref{ed:phi_z}), and $x^r$ can be obtained according to Proposition~\ref{ed:prop:2}.
For any $\ell_y(y)\in L^2(\mathbb Y)$, we define $h(x^r)$ as follows:
\begin{equation}\label{ed:def_g}
    h(x^r) \triangleq \ell_y(C'x^r) \overset{(\ref{ed:decompose_y})}{=} \ell_y(y)
\end{equation}

In the following, we show that $h(x^r)\in L^2(\mathbb X^r)$. First, consider the case when $C'$ is not a square. If Assumption~\ref{ed:ass:1} holds, $\ell_y (y)$ is a square-integrable real-valued function, and the following expression can be obtained~\cite{giaquinta2010mathematical}:
\begin{equation}\label{change_var}
    \int_{\mathbb Y} |\ell_y (y)|^2 dy = \int_{\mathbb{ X}^r} |\ell_y (C'x^r)|^2 \sqrt{\text{det}(C'^\top C')} dx^r  <\infty
\end{equation}
If Assumption~\ref{ed:ass:2} holds, $\sqrt{\text{det}(C'^\top C')}>0$. When $C'$ is a square matrix, $ \sqrt{\text{det}(C'^\top C')} $ in~(\ref{change_var}) reduces to $ |\text{det}(C')| $, and it holds that $|\text{det}(C')| > 0$. Consequently, we can conclude that
\begin{equation}\label{change_var_2}
    \Vert \ell_y(C'x^r) \Vert^2 = \int_{\mathbb{ X}^r} |\ell_y (C'x^r)|^2 dx^r<\infty
\end{equation}
According to~(\ref{ed:def_g}) and (\ref{change_var_2}), we have:
\begin{equation}\label{deepc:hxr}
        \Vert h(x^r) \Vert^2 =\int_{\mathbb{ X}^r} |h (x^r)|^2 dx^r
        \overset{(\ref{ed:def_g})}{=}\int_{\mathbb{ X}^r} |\ell_y (C'x^r)|^2 dx^r \overset{(\ref{change_var_2})}{<}\infty
\end{equation}
Since $\mathbf{\Phi}_r(x^r) =\{\phi_{i}^r(x^r)\}_{i=1}^\infty$ is a C.O.N.S. in $L^2(\mathbb X^r)$ from Proposition~\ref{ed:prop:1}, and $h(x^r)\in L^2(\mathbb X^r)$ as shown in~(\ref{deepc:hxr}), $h(x^r)$ can be represented in the following form:
\begin{equation}\label{ed:complete_1}
    h(x^r) = \sum_{i=1}^\infty a_i\phi_{i}^r(x^r)
\end{equation}
where $a_i = \langle h(x^r),\phi_{i}^r(x^r)\rangle$.

Therefore, it follows that
\begin{equation}
       \ell_y(y) \overset{(\ref{ed:def_g})}{=} h(x^r)  \overset{(\ref{ed:complete_1})}{=} \sum_{i=1}^\infty a_i\phi_{i}^r(x^r)
        \overset{(\ref{ed:phi_z})}{=} \sum_{i=1}^\infty a_i\phi_{i}^z(y)
\end{equation}
which indicates the completeness of $\mathbf{\Phi}_z(y)$. $\square$

\begin{theorem}\label{ed:thm:2}
If Assumptions~\ref{ed:ass:3}-\ref{ed:ass:1} hold, $\mathbf{\Phi}_z(y)$, of which the elements are defined in~(\ref{ed:phi_z}), and the C.O.N.S. $\mathbf{\Phi}_v(u)$ in $L^2(\mathbb U)$, as defined in Assumption~\ref{ed:ass:3}, satisfy both Condition 1 and Condition 2 stated in Section~\ref{sec:req}.
\end{theorem}
\textbf{Proof.} 
According to Proposition~\ref{ed:prop:1}, $\mathbf{\Phi}_z(y)$ is either a subsequence of, or is equivalent to $\mathbf{\Phi}(x)$. Therefore, $z = \Phi_z(y)$ in~(\ref{ed:model_B:2}) depends linearly on the lifted state $\Phi(x)$ in~(\ref{ed:model_B:1}) and can be described as follows:
\begin{equation}\label{ed:req_1:conclusion}
    z = \Phi_z(y) = E\Phi(x)
\end{equation}
where $\Phi(x)$ is constructed based on (\ref{ed:lifting function}); $E$ is a sparse matrix, the elements of which are 1 at positions corresponding to variables present in both $\Phi(x)$ and $\Phi_z(y)$, and 0 at positions corresponding to variables that appear only in $\Phi(x)$. Therefore, Condition~1 can be satisfied.

If Assumption~\ref{ed:ass:1} holds, since $\mathbf \Phi_z(y)$ is complete in $L^2(\mathbb Y)$ according to Proposition~\ref{ed:prop:3}, there exists a linear combination of the elements of the mapped output $z$ that can represent $\ell_y(y)$. Additionally, if Assumption~\ref{ed:ass:3} holds, then $\mathbf{\Phi}_{v}(u)$ is complete in $L^2(\mathbb U)$. Consequently, there exists a linear combination of the elements of the mapped input $v$ that can represent $\ell_u(u)$. Therefore, the linear terms (i.e., $P_zz_k$ and $P_vv_k$) on the right-hand side of (\ref{ed:cost_qua}) are sufficient to represent $\ell_e(y_k,u_k)$, which makes Condition~2 satisfied. $\square$

\subsection{Boundedness of the cost approximation error}
\label{sec:boundary}
From a practical perspective, finding infinite-dimensional mapped output ${\Phi}_z(y)$ and mapped input ${\Phi}_v(u)$ can be infeasible. To address this, we employ neural networks $F_\theta(\cdot)$ and $N_\gamma(\cdot)$ to find the finite-dimensional approximations of the mapped output ${\Phi}_z(y)$ and mapped input ${\Phi}_v(u)$. 
Let $n_z$ and $n_v$ be finite positive integers. If Assumptions~\ref{ed:ass:3}-\ref{ed:ass:1} hold, the finite-dimensional mapped output and mapped input vectors can be represented as
$z_{n_z} = \Phi_{z,n_z}(y)$ and $v_{n_v} = \Phi_{v,n_v}(u)$, where $\Phi_{z,n_z}(y)$ and $\Phi_{v,n_v}(u)$ are constructed from finite-dimensional output and input mappings $\mathbf \Phi_{z,n_z}(y)$ and $\mathbf \Phi_{v,n_v}(u)$, which consist of the first $n_z$ and $n_v$ components of $\mathbf \Phi_z(y)$ and $\mathbf \Phi_v(u)$, respectively.
In this finite-dimensional case, (\ref{ed:req_1:conclusion}) still holds since $\mathbf \Phi_{z,n_z}(y)$ is a subsequence of $\mathbf \Phi_z(y)$.
Meanwhile, the use of the finite-dimensional $\Phi_{z,n_z}(y)$ and $\Phi_{v,n_v}(u)$ will lead to approximation error for the economic cost, and in what follows, we analyze the boundedness of this approximate error.

The approximations of $l_y(y)$ and $l_u(u)$ using $\mathbf \Phi_{z,n_z}(y)$ and $\mathbf \Phi_{v,n_v}(u)$, denoted by $\ell_{y,n_z}(y)$ and $\ell_{u,n_v}(u)$, respectively, can be expressed as follows:
\begin{subequations}\label{ed:n-approx}
    \begin{align}
    \ell_{y,n_z}(y) &= \sum_{i=1}^{n_z} \langle \ell_{y}(y),  \phi_{i}^z(y)  \rangle \phi_{i}^z(y)\\
    \ell_{u,n_v}(u) &= \sum_{j=1}^{n_v} \langle \ell_{u}(u),  \phi_{j}^v(u)  \rangle \phi_{j}^v(u)
    \end{align}
\end{subequations}
The approximation errors of $\ell_y(y)$ and $\ell_u(u)$ in (\ref{ed:ecost}) caused by the use of finite-dimensional approximations, denoted as $E_{n_z}(y)$ and $E_{n_v}(u)$, are calculated as follows:
\begin{subequations}\label{ed:trun_e}
{\small
     \begin{align}
    E_{n_z}(y) &= \ell_y(y) - \ell_{y,n_z}(y) = \ell_y(y) - \sum_{i=1}^{n_z} \langle \ell_{y}(y),  \phi_{i}^z(y)  \rangle \phi_{i}^z(y)\label{ed:trun_e:y} \allowdisplaybreaks\\ 
    E_{n_v}(u) &= \ell_u(u) - \ell_{u,n_v}(u) = \ell_u(u) - \sum_{j=1}^{n_v} \langle \ell_{u}(u),  \phi_{j}^v(u)  \rangle \phi_{j}^v(u)\label{ed:trun_e:u}
     \end{align}
}
\end{subequations}
The approximation error of the economic cost function $\ell_e(y,u)$ in (\ref{ed:ecost}), denoted as $E(y,u)$,  is given by $E(y,u) = E_{n_z}(y) + E_{n_v}(u)$.
In this section, we show the boundedness and the convergence of $\Vert E(y,u)\Vert$.

\begin{theorem}\label{ed:thm:3}
    If Assumptions~\ref{ed:ass:3}-\ref{ed:ass:1} hold, $\Vert E(y,u) \Vert$ is bounded and converges to $0$ as $n_z$ and $n_v$ approach infinity.
\end{theorem}
\textbf{Proof.} 
If Assumptions~\ref{ed:ass:3}-\ref{ed:ass:1} hold, the output mapping $\mathbf \Phi_z(y)$, whose elements are defined in~(\ref{ed:phi_z}) can be constructed, and $\mathbf \Phi_{z,n_z}(y)$ is then obtained accordingly.
Let $c_i = \langle \ell_{y}(y),  \phi_{i}^z(y)\rangle$, $i=1,\dots, n_{z}$. If Assumption~\ref{ed:ass:1} is satisfied, based on the Cauchy-Schwarz inequality~\cite{steele2004cauchy}, it holds that:
\begin{subequations}\label{ed:ci_finite}
\begin{align}
        |c_i|^2 &= |\int_{\mathbb Y} \ell_y(y)\phi_{i}^z(y) dy | ^2
         \leq \int_{\mathbb Y} |\ell_y(y)|^2 dy \int_{\mathbb Y} |\phi_{i}^z(y)|^2 dy \\
         &= \int_{\mathbb Y} |\ell_y(y)|^2 dy < \infty
\end{align}
\end{subequations}
It follows that:
\begin{subequations}\label{ed:bound-1}
    \begin{align}
        &\Vert E_{n_z}(y) \Vert ^2 = \int_{\mathbb Y} |E_{n_z}(y)|^2dy  = \int_{\mathbb Y}    | \ell_y(y) - \sum_{i=1}^{n_z} c_i \phi_{i}^z(y)|^2    dy \label{ed:bound-1:1}\\
        & = \int_{\mathbb Y}   |\ell_y(y)|^2  - 2\ell_y(y) \sum_{i=1}^{n_z} c_i \phi_{i}^z(y)   + |\sum_{i=1}^{n_z}  c_i \phi_{i}^z(y)|^2   dy \label{ed:bound-1:3}
    \end{align}
\end{subequations}
We now show that all three components of~(\ref{ed:bound-1:3}) are integrable. From Assumption~\ref{ed:ass:1}, it follows that 
\begin{equation}\label{ed:bound-I1}
    \int_{\mathbb Y}   |\ell_y(y)|^2dy = \Vert \ell_y(y)\Vert ^2
\end{equation}
The integration of the second term in~(\ref{ed:bound-1:3}) is computed as follows:
\begin{subequations}\label{ed:bound-I2}
    \begin{align}
        &\int_{\mathbb Y} \ell_y(y) \sum_{i=1}^{n_z} c_i \phi_{i}^z(y) dy =  \sum_{i=1}^{n_z} c_i \int_{\mathbb Y}\ell_y(y) \phi_{i}^z(y)dy \label{ed:bound-I2:1} \allowdisplaybreaks\\
        & = \sum_{i=1}^{n_z} c_i \langle \ell_{y}(y),  \phi_{i}^z(y)\rangle =  \sum_{i=1}^{n_z} |c_i|^2 < \infty \label{ed:bound-I2:2}
    \end{align}
\end{subequations}
Based on the orthogonal property of $\{\phi_{i}^z(y)\}_{i=1}^{n_z}$, the integration of the third term in~(\ref{ed:bound-1:3}) can be expressed as follows:
\begin{subequations}\label{ed:bound-I3}
    \begin{align}
        &\int_{\mathbb Y} |\sum_{i=1}^{n_z}  c_i \phi_{i}^z(y)|^2   dy \\
        & = \int_{\mathbb Y } \big (\sum_{1 \leq i\leq {n_z}}| c_i \phi_{i}^z(y)|^2 + \sum_{\substack{1 \leq j,k \leq {n_z} \\ j \neq k}} 2 c_j c_k \phi_{j}^z(y)\phi_{k}^z(y)  \big) dy\label{ed:bound-I3:1} \\ 
        & = \sum_{1 \leq i\leq {n_z}} \int_{\mathbb Y} | c_i \phi_{i}^z(y)|^2 dy + \sum_{\substack{1 \leq j,k \leq {n_z} \\ j \neq k}} 2 \int_{\mathbb Y} c_j c_k \phi_{j}^z(y)\phi_{k}^z(y) dy \label{ed:bound-I3:2} \allowdisplaybreaks \\
        & = \sum_{i=1}^{n_z} \int_{\mathbb Y} | c_i \phi_{i}^z(y)|^2 dy = \sum_{i=1}^{n_z} | c_i| ^2 \int_{\mathbb Y} |\phi_{i}^z(y)|^2   dy = \sum_{i=1}^{n_z} | c_i| ^2\label{ed:bound-I3:3}
    \end{align}
\end{subequations}

According to~(\ref{ed:bound-I1}),~(\ref{ed:bound-I2}), and~(\ref{ed:bound-I3}),~(\ref{ed:bound-1}) is reformulated as:
\begin{subequations}\label{ed:bound-2}
    \begin{align}
        \Vert E_{n_z}(y) \Vert ^2 & = \int_{\mathbb Y}   |\ell_y(y)|^2 dy  - 2 \int_{\mathbb Y}\ell_y(y) \sum_{i=1}^{n_z} c_i \phi_{i}^z(y)  dy   \notag \\[-1.5ex]
    & + \int_{\mathbb Y} |\sum_{i=1}^{n_z}  c_i \phi_{i}^z(y)|^2   dy \label{ed:bound-2:1}  \\
        &=  \Vert \ell_y(y)\Vert ^2  - 2\sum_{i=1}^{n_z} | c_i| ^2 + \sum_{i=1}^{n_z} | c_i| ^2 \label{ed:bound-2:2}\\
        & = \Vert \ell_y(y)\Vert ^2  - \sum_{i=1}^{n_z} | c_i| ^2 < \Vert \ell_y(y)\Vert ^2\label{ed:bound-2:3}
    \end{align}
\end{subequations}
Therefore, $\Vert  E_{n_z}(y) \Vert ^2$, which depends on the dimension of the mapped output $n_z$, is upper bounded by the squared $L_2$-norm of $\ell_y(y)$. $\Vert E_{n_z}(y) \Vert$ is bounded by $\Vert \ell_y(y)\Vert$.

\begin{figure*}[t!]
    \centering
    \includegraphics[width=0.85\textwidth]{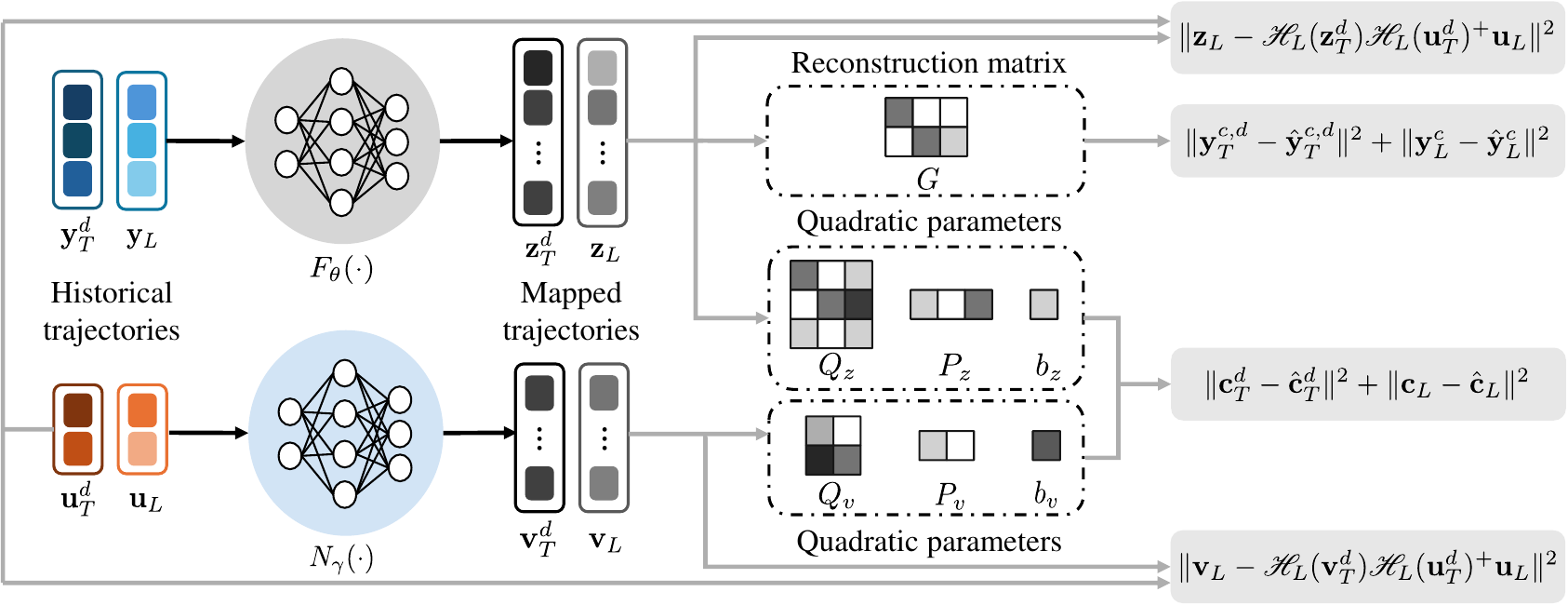}
    \caption{An overview of the training process for the proposed DeeEPC approach. }
    \label{ed:fig:train}
\end{figure*}

Moreover, by the Parseval's identity~\cite{muscat2024hilbert,cabrera1995hilbert}, the following equation holds:
\begin{equation}\label{ed:Parseval}
     \sum_{i=1}^\infty | c_i| ^2 = \Vert\ell_y(y)\Vert^2
\end{equation}
Considering the supremum in (\ref{ed:Parseval}), since $\sum_{i=1}^{n_z} | c_i| ^2$ is monotonically non-decreasing with respect to $n_z$, according to~\cite{bibby1974axiomatisations}, it will converge to $\Vert\ell_y(y)\Vert^2$ as $n_z \rightarrow \infty$. Consequently, the squared $L_2$-norm of $E_{n_z}(y)$ converges to $0$ as $n_z$ approaches infinity according to~(\ref{ed:bound-2:3}), and $\Vert E_{n_z}(y) \Vert$ also converges to 0.

Analogous conclusions can be drawn for the mapped input $v$ and the associated input-related cost function $\ell_u(u)$ by following the same line of reasoning as above. Therefore, $\Vert E(y,u) \Vert$, induced by the use of finite-dimensional $\Phi_{z,n_z}(y)$ and $\Phi_{v,n_v}(u)$, is bounded by $\Vert \ell_y(y)\Vert+\Vert \ell_u(u)\Vert$, and will converge to 0 as $n_z$ and $n_v$ approach infinity.
$\square$

\begin{remark}
The analysis of the boundedness of the economic cost approximation error in Theorem~\ref{ed:thm:3} may lead to slightly conservative results in the sense that it is conducted based on only the second and fifth terms on the right-hand side of (\ref{ed:cost_qua}), which account for the linear contributions of the mapped variables. When we apply the proposed approach, we choose to employ the full approximation form in (\ref{ed:cost_qua}), which also incorporates quadratic terms. Including these quadratic terms is expected to provide better approximation performance as compared to the case when only linear terms (i.e., the second and fifth terms on the right-hand side of (\ref{ed:cost_qua})) are used.
\end{remark}

\begin{remark}
Based on the theoretical results established in Section~\ref{sec:proof}, we identify a suitable choice of appropriate output and input mappings (denoted as $\Phi_z(y)$ and $\Phi_v(u)$ in the manuscript) that can generate $z$ and $v$. Specifically, $\Phi_z(y)$ and $\Phi_v(u)$ can be constructed from C.O.N.S. of the original output and the original input, respectively.
From an application point of view, finding C.O.N.S.s and constructing their corresponding finite-dimensional mappings as discussed in Section~\ref{sec:proof} can be challenging. Therefore, in this work, we choose to approximate the finite-dimensional input and output mappings, $\Phi_{z,n_z}(y)$ and $\Phi_{v,n_v}(u)$, by utilizing two neural networks and formulate the control optimization problems accordingly, as will be discussed in the next section.
\end{remark}

\section{Training and implementation}\label{sec:implementation}

\subsection{Training process}\label{sec:train}


Figure~\ref{ed:fig:train} illustrates the training process of the proposed \allowbreak DeeEPC. Two open-loop input and output trajectories of the system, denoted as $\mathbf{y}_T^d, \mathbf{u}_T^d$ and $\mathbf{y}_L, \mathbf{u}_L$, are collected. The training process involves learning the parameters of four key components, as detailed below:
\begin{itemize}
\item \textbf{Output mapping neural network $F_\theta(\cdot)$}:
    We employ a neural network $F_\theta(\cdot):\mathbb{R}^{n_y} \to \mathbb{R}^{n_z}$, with $\theta$ denoting its training parameters, to establish an $n_z$-dimensional approximation of the mapped output, i.e., $z_{n_z} = F_\theta(y)$.
    Two output trajectories $\mathbf y_T^d = \{y^d\}_1^T$ and $\mathbf y_L = \{y\}_1^L$ are reshaped into matrices $\mathbf{Y}_T^d = [y^d_1,\dots,y^d_T]^\top \in \mathbb{R}^{T\times n_y}$ and $\mathbf{Y}_L = [y_1,\dots,y_L]^\top\in \mathbb{R}^{L\times n_y}$, which serve as inputs to $F_\theta(\cdot)$. Two mapped output matrices $\mathbf{Z}^d_{T}\in \mathbb{R}^{T\times n_z}$ and $\mathbf{Z}_L\in \mathbb{R}^{L\times n_z}$ are generated as follows:
    \begin{equation}\label{deepc:nn1}
        \mathbf{Z}^d_{T} = F_\theta(\mathbf{Y}^d_T),\ \mathbf{Z}_L = F_\theta(\mathbf{Y}_L)
    \end{equation}
    From~(\ref{deepc:nn1}), the multi-step mapped output trajectories $\mathbf{z}^d_{T} = \{z_{n_z}^d\}_{1}^{T}$ and $\mathbf{z}_L = \{z_{n_z}\}_{1}^{L}$ can be extracted. 

\item \textbf{Input mapping neural network $N_\gamma(\cdot)$}:
    A neural network $N_\gamma(\cdot):\mathbb{R}^{n_u} \to \mathbb{R}^{n_v}$, where $\gamma$ represents the training parameters, is established to obtain the $n_v$-dimensional mapped input, i.e., $v_{n_v} = N_\gamma(u)$. Inputs to the neural network are matrices $\mathbf{U}_T^d = [u^d_1,\dots,u^d_T]^\top \in \mathbb{R}^{T\times n_u}$ and $\mathbf{U}_L = [u_1,\dots,u_L]^\top\in \mathbb{R}^{L\times n_u}$. $N_\gamma(\cdot)$ generates two mapped input matrices $\mathbf{V}^d_{T}\in \mathbb{R}^{T\times n_v}$ and $\mathbf{V}_L\in \mathbb{R}^{L\times n_v}$ as:
    \begin{equation}\label{deepc:nn2}
        \mathbf{V}^d_{T} = N_\gamma(\mathbf{U}^d_T),\ \mathbf{V}_L = N_\gamma(\mathbf{U}_L)
    \end{equation}
    with $\mathbf{V}^d_{T}$ and $\mathbf{V}_L$ contain information of multi-step mapped input trajectories, i.e., $\mathbf{v}^d_{T} = \{v_{n_v}^d\}_{1}^{T}$ and $\mathbf{v}_L = \{v_{n_v}\}_{1}^{L}$.

\item \textbf{Parameters of the quadratic cost function}: 
   The mapped outputs $z_{n_z}$ and mapped inputs $v_{n_v}$ generated by neural networks $F_\theta(\cdot)$ and $N_\gamma(\cdot)$ enable the construction of a \allowbreak quadratic economic cost function $\hat{\ell}_e(z_{n_z},v_{n_v})$ following~(\ref{ed:cost_qua}). Specifically, we approximate the economic cost at sampling instant $k$, denoted by $\hat c_k$, as follows:
    \begin{equation}\label{deepc:c}
        \hat{c}_k = z_{n_z,k}^\top Q_zz_{n_z,k} + P_zz_{n_z,k} +b_z + v_{n_v,k}^\top Q_vv_{n_v,k} + P_vv_{n_v,k} +b_v
    \end{equation}
    A similar quadratic formulation is applied to $z_{n_z}^d$ and $v_{n_v}^d$ in the same manner as~(\ref{deepc:c}).
    To ensure the symmetry and positive definiteness of $Q_z$ and $Q_v$, we parameterize the two matrices as $Q_z = \text{diag}\big(\text{exp}(q_z)\big)$ and $Q_v = \text{diag}\big(\text{exp}(q_v)\big)$, where $q_z\in\mathbb{R}^{n_z}$ and $q_v\in\mathbb{R}^{n_v}$ are real-valued vectors. The parameters $q_z$, $q_v$, $P_z$, $P_v$, $b_z$ and $b_v$ are trainable parameters.

\item \textbf{Output reconstruction matrix}: 
    Let $y^c\in \mathbb Y_c \subseteq \mathbb R^{n_c}$ denote the subset of output variables on which hard constraints need to be enforced. The trainable reconstruction matrix $G\in\mathbb{R}^{n_c\times n_z}$ maps the latent output $z_{n_z,k}$ to an approximation of $y^c$.
    This mapping will be used for the enforcement of system constraints during online implementation. The reconstructed output at sampling instant $k$ is given by:
    \begin{equation}\label{deepc:re}
        \hat{y}_k^c = Gz_{n_z,k}
    \end{equation}
    where $\hat{y}_k^c$ is an approximation of the output vector $y^c_k$.
\end{itemize}

During the training stage, a dataset $\mathcal{D}:=\{\mathbf{u}^d_T,\mathbf{u}_L,\mathbf{y}^d_T,\mathbf{y}_L, \mathbf{c}^d_T, \allowbreak \mathbf{c}_L\}$ is collected from open-loop process operation, where $\mathbf{c}_T^d:=\{c^d\}_1^{T}$ and $ \mathbf{c}_L:=\{c\}_1^{L}$ represent the sequences of economic cost values associated with $\mathbf{y}^d_T, \mathbf{u}^d_T$ and $\mathbf{y}_L,\mathbf{u}_L$, respectively. The objective is to optimize the trainable parameters $\theta$, $\gamma$, $q_z$, $q_v$, $P_z$, $P_v$, $b_z$, $b_v$, and $G$. 
The optimization problem for training is formulated as follows:
\begin{equation}\label{deepc:op_train}
        \min_{\theta, q,P,b,G}\ \mathcal{L} = \min_{\theta, q,P,b,G}\  \alpha_1 \mathcal{L}_{e} + \alpha_2 \mathcal{L}_{re}  + \alpha_3  \mathcal{L}_{z} + \alpha_4  \mathcal{L}_{v}
\end{equation}
where $\mathcal{L}_{e}$ denotes the loss for economic cost approximation; $\mathcal{L}_{re}$ denotes the reconstruction loss of the set of output variables that need to satisfy hard constraints; $\mathcal{L}_{z}$ and $\mathcal{L}_{v}$ are the loss terms formulated based on Willems' fundamental lemma~\cite{1428856}; $\alpha_i,i=1,2,3,4$, are user-specified weighting coefficients. These weights are chosen to normalize the magnitudes of the four loss terms in (\ref{deepc:op_train}) at the start point of the training process to prevent any single loss term from dominating the gradient calculation.

The first term $\mathcal{L}_{e}$ on the right-hand side of~(\ref{deepc:op_train}) quantifies the representation error of the economic cost, and is given as follows:
\begin{equation}
\label{deepc:lcost}
    \mathcal{L}_{e} = \mathbb{E}_{\mathcal{D}}\ \left(||\mathbf{c}_T^d - \hat{\mathbf{c}}_T^d||^2 + ||\mathbf{c}_L - \hat{\mathbf{c}}_L||^2\right)
\end{equation}
where $ \hat{\mathbf{c}}_T^d:=\{\hat c^d\}_1^{T}$ and $ \hat{\mathbf{c}}_L:=\{\hat c\}_1^{L}$ are the approximated economic cost trajectories obtained from the corresponding mapped output sequences $\mathbf{z}^d_T$, $\mathbf{z}_L$, and mapped input sequences $\mathbf{v}^d_T$, $\mathbf{v}_L$, as defined in~(\ref{deepc:c}).

The second term $\mathcal{L}_{re}$ on the right-hand side of~(\ref{deepc:op_train}) is for the discrepancy between the ground-truth and the reconstructed outputs, as defined below:
\begin{equation}
\label{deepc:lstate}
\mathcal{L}_{re} = \mathbb{E}_{\mathcal{D}}\  \left(\big|\big| \mathbf{y}^{c,d}_T - \hat{\mathbf{y}}^{c,d}_T \big|\big|^2+ \big|\big|  \mathbf{y}^{c}_L - \hat{\mathbf{y}}^{c}_L   \big|\big|^2\right)
\end{equation}
where $\mathbf{y}^{c,d}_T:=\{y^{c,d}\}_0^{T-1}$ and $\mathbf{y}^{c}_L:=\{y^{c}\}_0^{L-1}$ contain the subset of variables from $\mathbf{y}_T^d$ and $\mathbf{y}_L$ on which hard constraints need to be imposed; $\hat{\mathbf{y}}^{c,d}_T:=\{\hat y^{c,d}\}_0^{T-1}$ and $\hat{\mathbf{y}}^{c}_L:=\{\hat y^{c}\}_0^{L-1}$ represent the reconstructed output calculated based on~(\ref{deepc:re}). 

The third and fourth terms $\mathcal{L}_{z}$ and $\mathcal{L}_{v}$ on the right-hand side of~(\ref{deepc:op_train}), derived from Willems' fundamental lemma~\cite{1428856}, are defined as follows:
\begin{subequations}
\label{deepc:lg}
\begin{align}
\mathcal{L}_{z} &= \mathbb{E}_{\mathcal{D}}\ ||\mathbf{z}_L- \mathscr{H}_L(\mathbf{z}_T^d)\mathscr{H}_{L}(\mathbf{u}_T^d)^+\mathbf{u}_L||^2 \\
\mathcal{L}_{v} &= \mathbb{E}_{\mathcal{D}}\ ||\mathbf{v}_L- \mathscr{H}_L(\mathbf{v}_T^d)\mathscr{H}_{L}(\mathbf{u}_T^d)^+\mathbf{u}_L||^2 
\end{align}
\end{subequations}
where $\mathscr{H}_{L}(\mathbf{u}_T^d)^+$ is the pseudo-inverse of $\mathscr{H}_{L}(\mathbf{u}_T^d)$.
Note that $\mathscr{H}_{L}(\mathbf{u}_T^d)^+u = \mathscr{H}_{L}(\mathbf{u}_T^d)^\top (\mathscr{H}_{L}(\mathbf{u}_T^d)\mathscr{H}_{L}(\mathbf{u}_T^d)^\top)^{-1} u $ is one of the feasible solutions with the minimum $L_2$ norm. 
It describes the dynamic behaviors of the mapped output and the mapped input, as shown in~(\ref{deepc:lg}).

\subsection{Online implementation} \label{sec:control}

After training, the optimal parameters, denoted by $\theta^*$, $\gamma^*$, $Q_z^*$, $Q_v^*$, $P_z^*$, $P_v^*$, $b_z^*$, $b_v^*$, and $G^*$, are used for online control implementation of the proposed DeeEPC. 

To address mismatch and unknown disturbances, we adopt a regularized DeeEPC formulation, based on~\cite{coulson2019data}, as below:
\begin{subequations}\label{deepc:reg economic DeePC_opt}
\begin{align}
    \min_{\bar g_k,\ \mathbf\sigma_{z,k} }\sum_{j=k}^{k+N_p-1}&\lambda(\hat z_{n_z,j|k}^\top Q_z^*\hat z_{n_z,j|k} + P_z^*\hat z_{n_z,j|k}+b_z^* + \hat v_{n_v,j|k}^\top Q_v^*\hat v_{n_v,j|k} + \notag \\[-2.2ex]
    & P_v^*\hat v_{n_v,j|k}+b_v^*) + \Vert \Delta \hat{u}_{j|k} \Vert_R^2 + \Vert\mathbf\sigma_{z,k} \Vert_{\beta_z}^2 + \Vert \bar g_k \Vert_{\beta_g}^2 \label{deepc:reg economic DeePC_opt:prob}\\
    \text{s.t.}\ &\left[\begin{array}{c}
            \bar U_p \\
            \bar V_p \\
            \bar Z_p \\
            \bar U_f \\
            \bar V_f \\
            \bar Z_f\\
          \end{array}
    \right]\ \bar g_k = 
    \left[\begin{array}{c}
         \mathbf u_{ini,k} \\
         \mathbf v_{ini,k} \\
         \mathbf z_{ini,n_z,k} \\
         \hat{\mathbf{u}}_k \\
         \hat{\mathbf{v}}_k \\
         \hat{\mathbf{z}}_{n_z,k}
    \end{array} 
    \right] + 
    \left[\begin{array}{c}
         \mathbf 0 \\
         \mathbf 0 \\
         \mathbf \sigma_{z,k} \\
         \mathbf 0 \\
         \mathbf 0 \\
         \mathbf 0
    \end{array} 
    \right] 
    \label{deepc:reg economic DeePC_opt:1}\\
    &\hat{u}_{j|k}\in\mathbb{U},\ j = k,\dots,k+N_p-1\label{deepc:reg economic DeePC_opt:2} \\
    &G^*\hat{z}_{n_z,j|k} \in \mathbb{Y}_c,\ j = k,\dots,k+N_p-1\label{deepc:reg economic DeePC_opt:3}
\end{align}
\end{subequations}
where $R\in\mathbb{R}^{n_u\times n_u}$ is a positive-definite matrix;
$\Delta \hat u_{j|k}:= \hat u_{j|k} - \hat u_{j-1|k}$ is the rate of change in the control input, where $\hat u_{k-1|k}$ corresponds to the last element of $\mathbf u_{ini,k}$; $\mathbf \sigma_{z,k} \in \mathbb{R}^{T_{ini}\times n_z}$ is a slack variable that accounts for mismatch and unknown disturbances.

\begin{algorithm}[t!]
\caption{Online Implementation of the Proposed DeeEPC}\label{ed:reg_edeepc_alg}
\begin{algorithmic}[1]

\vspace{6pt}
\Input Offline collected input-output trajectories $\mathbf{u}_T^d$, $\mathbf{y}_T^d$; 
optimal parameters $\theta^*$, $\gamma^*$, $Q_z^*$, $Q_v^*$, $P_z^*$, $P_v^*$, $b_z^*$, $b_v^*$, $G^*$; 
weighting parameters $\lambda$, $R$, $\beta_z$, and $\beta_g$
\vspace{3pt}
\Output Control input $u_k \in \mathbb{U} \subseteq \mathbb{R}^{n_u}$

\vspace{6pt}
\State Set $k=0$
\For{$k = 0$ to $T_{\text{ini}} - 1$}
    \State Collect $T_{\text{ini}}$-step initialization trajectories
\EndFor
\State \Return $\mathbf{y}_{ini, T_{ini}}$, $\mathbf{u}_{ini, T_{ini}}$

\vspace{6pt}
\State Compute $\mathbf{z}_T^d$ and $\mathbf{z}_{ini, T_{ini}}$ following~(\ref{deepc:nn1})
\State Compute $\mathbf{v}_T^d$ and $\mathbf{v}_{ini, T_{ini}}$ following~(\ref{deepc:nn2})
\State Construct Hankel matrices $[\bar{U}_p, \bar{V}_p, \bar{Z}_p, \bar{U}_f, \bar{V}_f, \bar{Z}_f]^\top$

\vspace{6pt}
\While{$k \geq T_{\text{ini}}$}
    \State Solve~(\ref{deepc:reg economic DeePC_opt}) to obtain $\bar{g}_k^*$
    \State Compute $\hat{\mathbf{u}}_k^*$ following~(\ref{deepc:optu})
    \State Apply control input $u_k = \hat{u}_{k|k}^*$ to system~(\ref{ed:nlmodel}), observe $y_k$ and $c_k$
    \State Compute latest $z_k$ and $v_k$ using~(\ref{deepc:nn1}) and (\ref{deepc:nn2})
    \State Update $\mathbf u_{ini,k+1}$, $\mathbf z_{ini,k+1}$, and $\mathbf v_{ini,k+1}$
    \State $k \leftarrow k + 1$
\EndWhile

\vspace{6pt}
\end{algorithmic}
\end{algorithm}

To execute~(\ref{deepc:reg economic DeePC_opt}), as presented in Figure~\ref{ed:fig:control}, the mapped output and mapped input trajectories $\mathbf z_{T,n_z}^d:=\{z_{n_z}^d\}_{1}^T$ and $\mathbf v_{T,n_v}^d:=\{v_{n_v}^d\}_{1}^T$ and the initialization trajectories $\mathbf z_{ini,n_z,k}:=\{z_{n_z}\}_{k-T_{ini}}^{k-1}$ and $\mathbf v_{ini,n_v,k}:=\{v_{n_v}\}_{k-T_{ini}}^{k-1}$ are generated by the trained neural networks $F_{\theta^*}(\cdot)$ and $N_{\gamma^*}(\cdot)$. $\hat{\mathbf{z}}_{n_z,k}:=  \{\hat{z}_{n_z}\}_{k|k}^{k+N_p-1|k}$ and $\hat{\mathbf{v}}_{n_v,k}:=  \{\hat{v}_{n_v}\}_{k|k}^{k+N_p-1|k}$ are the predictions of mapped output and input sequences, respectively.

The proposed control optimization problem in (\ref{deepc:reg economic DeePC_opt}) can remain computationally tractable even when complex neural network structures are employed, since the neural networks $F_\theta(\cdot)$ and $N_\gamma(\cdot)$ are not directly involved in the formulation or solution of the optimization problem. Moreover, the number of rows in the Hankel matrices in (\ref{deepc:reg economic DeePC_opt:1}) remains manageable for various large-scale systems, since the implementation of the proposed method relies on the system inputs and outputs rather than the full system state.
The dimension of the operator $g$ depends on the column number of Hankel matrices, and can be large. We employ the singular value decomposition (SVD)-based dimension reduction method in~\cite{zhang2023dimension} to reduce the computational complexity. In~(\ref{deepc:reg economic DeePC_opt:1}), $\bar U_p, \bar V_p, \bar Z_p, \bar U_f, \bar V_f, \bar Z_f $ are partitions of the reduced-order Hankel matrices. $\bar g_k$ is the corresponding reduced-order operator of dimension $n_g$. $\beta_z\in\mathbb R^{n_zN_p\times n_zN_p}$ and $\beta_g\in\mathbb R^{n_gN_p\times n_gN_p}$ are weight matrices for the two regularization terms. 

The optimal input sequence $\hat{\mathbf{u}}_k^*$ is computed according to the optimal operator $\bar g_k^*$ as follows:
\begin{equation}\label{deepc:optu}
    \hat{\mathbf{u}}_k^* = U_f\bar g_k^*
\end{equation}
where $\hat{\mathbf{u}}_k^* = \big[\hat{u}^{* \top}_{k|k}, \ldots, \hat{u}_{k+N_p-1|k}^{* \top} \big]^\top$. The first element $\hat{u}^*_{k|k}$ is applied to the nonlinear system~(\ref{ed:nlmodel}). The online implementation of the regularized DeeEPC is summarized in Algorithm \ref{ed:reg_edeepc_alg}.

\section{Case study on the wastewater treatment plant}\label{sec:wwtp}

\begin{figure}[t]
    \centering
    \includegraphics[width=0.48\textwidth]{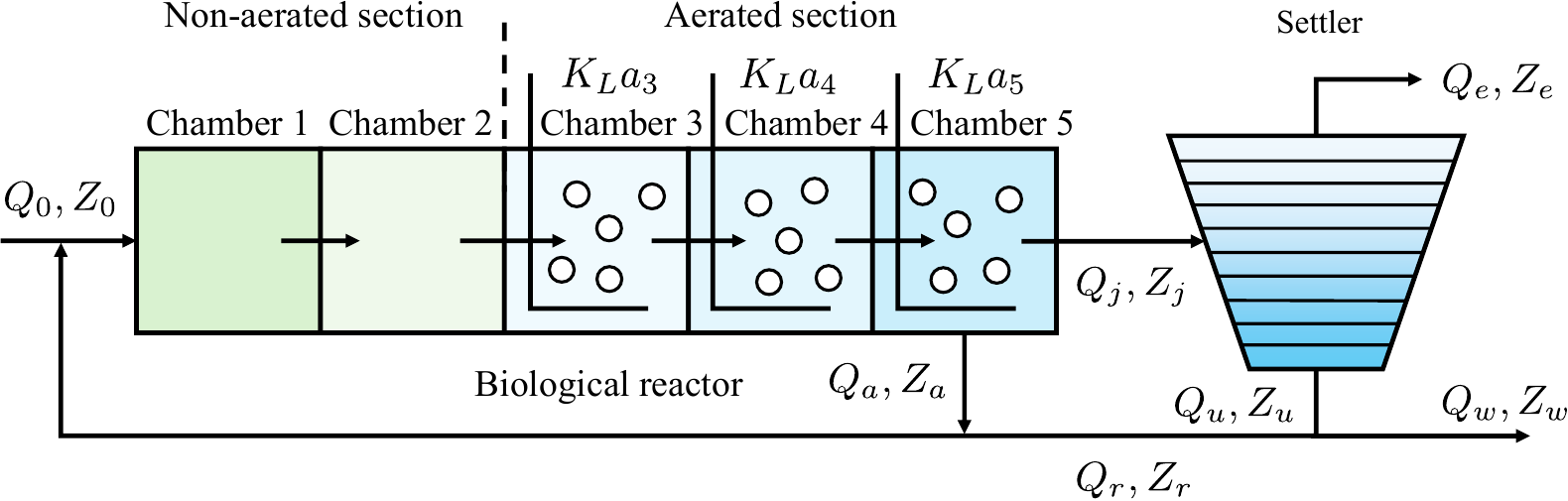}
    \caption{A schematic of biological wastewater treatment process~\cite{alex2008benchmark}.}
    \label{fig:wwtp}
\end{figure}
\subsection{Process description}
First, we consider a biological wastewater treatment plant, of which the dynamics are described by the Benchmark Simulation Model No.1 (BSM1)~\cite{alex2008benchmark}. 
This plant contains a five-chamber biological reactor and a secondary settler. Figure~\ref{fig:wwtp} presents a schematic of this plant. 

Among the five chambers of the biological reactor, the first two constitute the anoxic section, and the remaining three chambers account for the aerated section. Denitrification process takes place in the first two chambers of the biological reactor, where bacteria convert nitrates into nitrogen. The remaining three chambers facilitate the nitrification process, where bacteria oxidize ammonium to nitrates. In total, 145 state variables are considered, and 145 ordinary differential equations (ODEs) are used to describe the dynamic behaviors of this plant. A comprehensive description of the process and the values of the key parameters can be found in~\cite{alex2008benchmark}.

The real-time economic cost is calculated as a weighted sum of the effluent quality and overall operational cost as follows~\cite{alex2008benchmark,zeng2015economic}:
\begin{equation}\label{deepc:wwtp:cost}
    c_k = \text{EQ}_k + 0.3\text{OCI}_k
\end{equation}
where $\text{EQ}_k$ measures the effluent quality and $\text{OCI}_k$ represents the overall cost index at time instant $k$.

\subsection{Settings}\label{section:settings}

\begin{table}[t]
\renewcommand\arraystretch{1.25}
\caption{Hyperparameters used in the training of the proposed DeeEPC.} \label{table:trainingparameter}\vspace{2mm}
\centering \small
\begin{tabular}{ c c }
      \toprule
      Parameters& Values \\
      \midrule
      Batch size & 128\\
      Structure of $F_\theta(\cdot)$ and $N_\gamma(\cdot)$ & (128, 256) \\
      Training epoch & 100\\
      Activate function & ReLU\\
       Optimizer & Adam~\cite{kingma2014adam}\\
      Learning rate for $F_\theta(\cdot)$ and $N_\gamma(\cdot)$ & $10^{-4}$\\
      Learning rate for $Q,P,b,G$ & $10^{-3}$\\
      \bottomrule
\end{tabular}
\end{table}

We design controllers for the wastewater treatment plant (WWTP) based on four predictive control methods, including the proposed regularized DeeEPC approach as described in~(\ref{deepc:reg economic DeePC_opt}), the set-point tracking nonlinear MPC method in~\cite{zeng2015economic}, nonlinear EMPC based on the first-principles model in~\cite{zeng2015economic} (referred to as nonlinear EMPC), and learning-based input-output Koopman EMPC in~\cite{han2024efficient}. In addition, we consider a convex DeePC method without nonlinear mappings (referred to as convex DeePC). Specifically, the original economic cost function is approximated by a quadratic function of the original output and the original input, given by $\bar \ell_e(y_k,u_k) = y_k^\top Q_y y_k + P_y y_k + b_y + u_k^\top Q_u u_k + P_uu_k + b_u$, where $\bar \ell_e(y_k,u_k)$ denotes the approximated economic cost function adopted in the convex DeePC formulation.

The hyperparameters of the proposed regularized DeeEPC are as follows: $T = 10^3$, $T_{ini} = 2$, $N_p = 2$, $\beta_z=5\times 10^{10}$, $\beta_g=1\times 10^{-8}$, $n_z = 60$, and $n_v = 4$. The dimensions $n_z$ and $n_v$ are determined through trial-and-error simulations, with the aim of optimizing the training performance. The regularization parameters in (\ref{deepc:reg economic DeePC_opt:prob}) are chosen as $\beta_z = 5\times 10^{10}$ and $\beta_g = 1\times 10^{-8}$. The hyperparameters for training are shown in Table~\ref{table:trainingparameter}. A PID controller~\cite{seborg2016process} is designed to initialize the regularized DeeEPC, that is, to generate and compute initialization trajectories in~(\ref{deepc:reg economic DeePC_opt}) at the initial time instant. The objective of PID control is to track a set-point $y_{s} = [S_{NO,2}^{s}, S_{O,5}^{s}]^\top = [1\ (\text{gN/m}^3), 2\ \big(\text{g(-COD/m)}^3\big)]^\top$ considered in~\cite{zeng2015economic}. 
The weighting matrices of the rate of change in input $\Vert \Delta u \Vert$ for the set-point tracking MPC and EMPC based on~\cite{zeng2015economic} are both $0.001I_2$. 
The weighting matrix for output deviations is made the same as those adopted in~\cite{zeng2015economic}. 
The control horizon of the tracking MPC and nonlinear EMPC is set to be the same as that of the proposed regularized DeeEPC.
As compared to~\cite{zeng2015economic}, the objective function of nonlinear EMPC based on~\cite{zeng2015economic} is modified by incorporating a penalty on the rate of change in the control input and excluding the terminal cost. 
For the input-output Koopman EMPC method~\cite{han2024efficient} (referred to as Koopman EMPC), the learning rate for model training is set as $10^{-5}$, and the trainable parameters are trained over 100 epochs. The other parameters, including neural network structure, the dimension of the lifted state, and the control horizon, are consistent with those in~\cite{han2024efficient}.
The control objective of the proposed DeeEPC, the nonlinear EMPC based on~\cite{zeng2015economic}, and the Koopman EMPC based on~\cite{han2024efficient} is to minimize the overall economic cost $c_k$ defined in~(\ref{deepc:wwtp:cost}).
The objective of the nonlinear MPC is to track the set-point $y_{s} = [S_{NO,2}^{s}, S_{O,5}^{s}]^\top = [1\ (\text{gN/m}^3), 2\ \big(\text{g(-COD/m)}^3\big)]^\top$, which is consistent with that in~\cite{zeng2015economic}.

\subsection{Data generation}
Input and output trajectories for Hankel matrix construction and neural network training are obtained from open-loop process simulations. Three different weather conditions are considered: dry, rainy, and stormy conditions, with the corresponding data available in~\cite{waterweb}. We collect open-loop data for neural network training under the dry weather condition. The proposed regularized DeeEPC controller is then tested under three weather conditions to evaluate its robustness against unseen disturbances.

Following the design in~\cite{alex2008benchmark}, the system input variables include the rate of the recycle flow and the oxygen transfer coefficient in the fifth chamber, that is, $u=[Q_a, K_La_5]^\top$. The system output variable $y\in \mathbb R^{43}$ includes 41 state variables used to calculate the economic cost, as well as 2 variables $y^c = [S_{NO,2}, S_{O,5}]^\top$ on which hard constraints should be imposed. 
The input constraints are shown in Table~\ref{table:inputbound}. The output constraints imposed on $y^c$ are provided in Table~\ref{table:constraints}.

The sampling time interval is $\Delta= 15\ \text{min}$. The initial condition of the state variables $x_0$ is adopted from~\cite{zeng2015economic}. Stochastic disturbances added to the process are generated following the Gaussian distribution $\mathcal{N}(0, 10^{-6}x_0^2)$ and then made bounded within a range of $[-0.1x_0, 0.1x_0]$. The control input $u$ trajectories for open-loop simulations are generated randomly within the defined boundaries. Each sampled value of the system input holds for 5 hours with an additional Gaussian noise $\mathcal{N}(0, 10^{-4}u_s^2)$, where $u_s = [1.4\times10^2~(\text{m}^3/\text{day}), 2.0\times10^4~(\text{day}^{-1})]^\top$.

\begin{table}[t]
  \renewcommand\arraystretch{1.25}
  \caption{The input constraints of the wastewater treatment plant~\cite{zeng2015economic}.}\label{table:inputbound}\vspace{2mm}
  \centering\small
    \begin{tabular}{ c c c}
      \toprule
       & lower bound & upper bound \\
      \midrule
      $Q_a$ ($\text{m}^3/\text{day}$)& 0  & 92230\\
      $K_La_5$ ($\text{day}^{-1}$)& 0  & 240 \\
      \bottomrule
    \end{tabular}
\end{table}

\begin{table}[t]
  \renewcommand\arraystretch{1.25}
  \caption{The output constraints of the wastewater treatment plant~\cite{zeng2015economic}.}\label{table:constraints}\vspace{2mm}
  \centering\small
    \begin{tabular}{ c c c}
      \toprule
       & lower bound & upper bound \\
      \midrule
      $S_{NO,2}$ ($\text{gN}/\text{m}^3$)& 0  & 10 \\
      $S_{O,5}$ ($\text{g(-COD)}/\text{m}^3$)& 0  & 10 \\
      \bottomrule
    \end{tabular}
\end{table}

When applying the proposed regularized DeeEPC controller, $4.5\times 10^3$ input and output data points are collected via open-loop simulations for training: $10^3$ data points are used for constructing Hankel matrices; $3.5\times 10^3$ data points are used for the training of the proposed regularized DeeEPC. When applying the learning-based input-output Koopman EMPC controller in~\cite{han2024efficient}, $5.4\times 10^4$ samples are used for training.

\subsection{Control performance}

\begin{figure}[!t]
    \centering
    \includegraphics[width=0.48\textwidth]{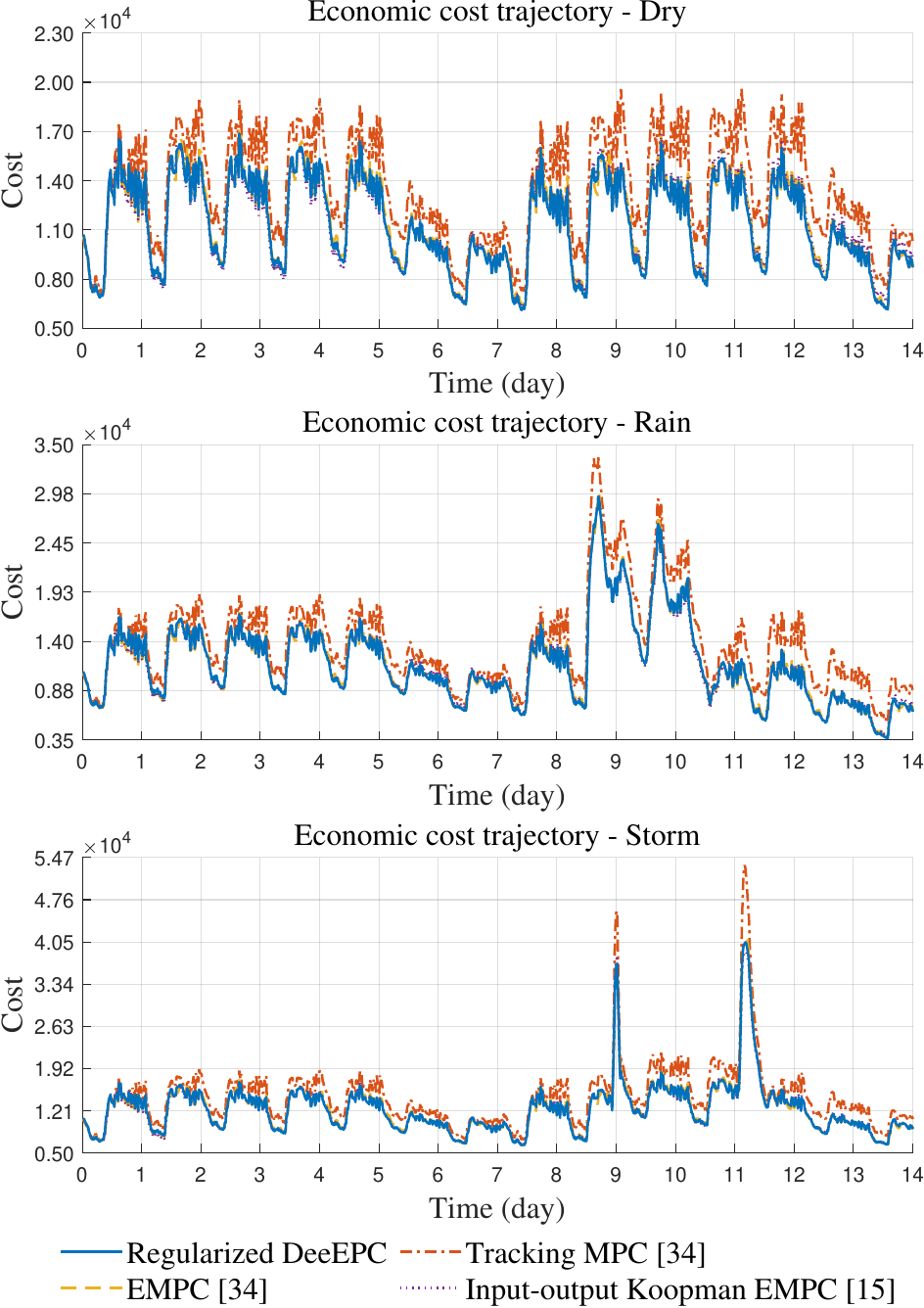}
    \caption{The economic cost trajectories of the wastewater treatment system under three weather conditions.}
    \label{fig:wwtp_cost}
\end{figure}

The optimization problem in (\ref{deepc:reg economic DeePC_opt}) is solved using Interior Point OPTimizer (IPOPT)~\cite{wachter2006implementation}. 
The trajectories of the economic cost under three different weather conditions are shown in Figure~\ref{fig:wwtp_cost}. 
As shown in Figure~\ref{fig:wwtp_cost}, the economic control approaches provide comparable performance and significantly outperform the set-point tracking MPC in terms of economic performance. Among them, the proposed regularized DeeEPC consistently achieves the lowest economic cost for most of the time horizon. To provide a quantitative comparison, the average economic costs under three weather conditions are reported in Table~\ref{table:avgcost}, where the proposed method provides the lowest value under all three weather conditions.
All the economic costs presented in Figure~\ref{fig:wwtp_cost} and Table~\ref{table:avgcost} are evaluated based on the economic cost function of the simulated wastewater treatment plant in (\ref{deepc:wwtp:cost}).
The proposed regularized DeeEPC achieves a reduction in the average economic cost of 15.46\%–16.29\% compared with set-point tracking MPC, 0.60\%–1.62\% compared with EMPC, 0.46\%–0.94\% compared with Koopman EMPC, and 1.95\%–2.86\% compared with the convex DeePC.

\begin{table}[!htb]
  \renewcommand\arraystretch{1.25}
  \caption{The average economic cost results of the wastewater treatment plant under different weather conditions.}\label{table:avgcost}\vspace{2mm}
  \centering 
  {\fontsize{8.4pt}{12pt}\selectfont
    \begin{tabular}{>{\centering\arraybackslash}p{2.6cm} 
                >{\centering\arraybackslash}p{1.5cm} 
                >{\centering\arraybackslash}p{1.5cm}
                >{\centering\arraybackslash}p{1.5cm}}  
      \toprule
       & Dry & Rain & Stormy \\
      \midrule
      \rule{0pt}{2.5ex} Regularized DeeEPC& $1.1359 \times 10^4$ &$1.1667 \times 10^4$  &$1.2328 \times 10^4$ \\
      \rule{0pt}{2.5ex} Convex DeePC& $1.1694 \times 10^4$ & $1.1933 \times 10^4$  &$1.2573 \times 10^4$ \\
      \rule{0pt}{2.5ex} Tracking MPC~\cite{zeng2015economic} & $1.3546\times 10^4$ & $1.3938\times 10^4$& $1.4582\times 10^4$\\
      \rule{0pt}{2.5ex} EMPC~\cite{zeng2015economic}& $1.1546\times 10^4$ & $1.1737\times 10^4$ & $1.2522\times 10^4$ \\
      Koopman EMPC~\cite{han2024efficient}
      & $1.1467\times 10^4$ & $1.1721\times 10^4$ & $1.2438\times 10^4$ \\
      \bottomrule
    \end{tabular}
    }
\end{table}

\begin{figure}[!t]
    \centering
    \includegraphics[width=0.48\textwidth]{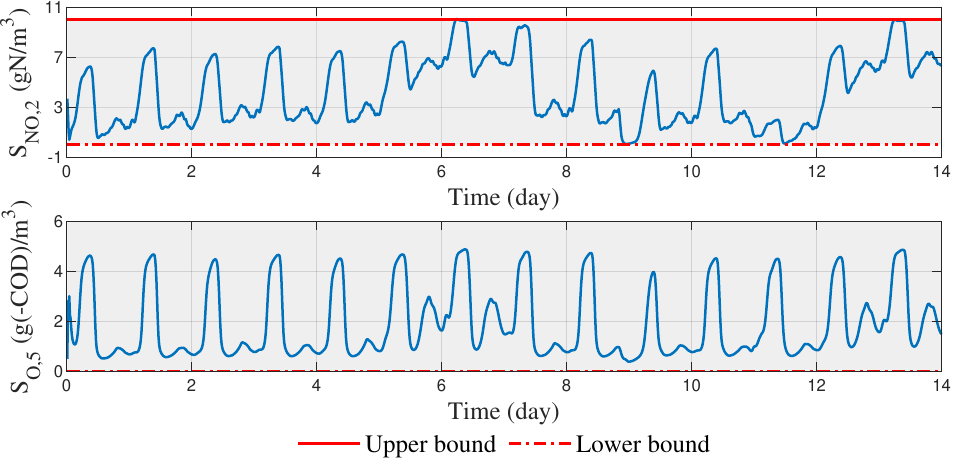}
    \caption{The $y^c$ trajectories of the wastewater treatment plant under the dry weather condition.}
    \label{fig:output_wwtp}
\end{figure}

Additionally, we assess whether $y^c$ satisfies the hard constraints in Table~\ref{table:constraints}. The trajectories of the constrained system outputs under dry weather are shown in Figure~\ref{fig:output_wwtp}. These trajectories remain within the shallow region for most of the time, such that the hard constraints are satisfactorily met. Under the other two weather conditions, the constraints are also met for the majority of the time. This demonstrates that the proposed approach can reliably enforce hard output constraints with the well-trained output mapping neural network $F_\theta(\cdot)$ and output reconstruction matrix $G$.

We also evaluate the computational efficiency of the proposed approach. 
The proposed method solves a convex quadratic programming (QP) problem at each time instant. The optimization problem associated with the proposed DeeEPC can be efficiently solved in polynomial time~\cite{floudas1995quadratic, potra2000interior, huang2021decentralized}.
The average computational times of the proposed method, the nonlinear EMPC in \cite{zeng2015economic}, and the convex DeePC are shown in Table \ref{table:avgtime}. 
On average, the proposed approach reduces the average computation time by 95.79\% as compared to the EMPC across the three weather conditions.
The convex DeePC formulation without lifted variables is more computationally efficient than the proposed method. Meanwhile, in practical applications, economic operational performance is typically prioritized over computation time, particularly when both methods exhibit fast online computation. Therefore, in view of its improved economic performance and comparable computational efficiency, the proposed DeeEPC method demonstrates superior overall performance as compared to the convex DeePC.

\begin{table}[!tb]
  \renewcommand\arraystretch{1.25}
  \caption{Average computational time (in seconds) of the proposed Regularized DeeEPC and EMPC~\cite{zeng2015economic} under different weather conditions.}\label{table:avgtime}\vspace{2mm}
  \centering 
  {\fontsize{8.4pt}{12pt}\selectfont
    \begin{tabular}{>{\centering\arraybackslash}p{2.6cm} 
                >{\centering\arraybackslash}p{1.01cm} 
                >{\centering\arraybackslash}p{1.01cm}
                >{\centering\arraybackslash}p{1.01cm}
                >{\centering\arraybackslash}p{1.03cm}}  
      \toprule
       & Dry & Rain & Stormy & Average \\
      \midrule
      \rule{0pt}{2.5ex} Regularized DeeEPC & 0.42384  & 0.57467   &0.60394 & 0.53415  \\
      \rule{0pt}{2.5ex} Convex DeePC & 0.25260  & 0.25081   & 0.28639 & 0.26327  \\
      \rule{0pt}{2.5ex} EMPC~\cite{zeng2015economic}& 11.7650  &  10.9122 &  15.3449 & 12.6740  \\
      \bottomrule
    \end{tabular}
    }
\end{table}

\begin{table}[thbp]
  \renewcommand\arraystretch{1.25}
  \caption{The average economic cost results under dry weather using various configurations.}\label{table:sensiticity}
  \centering\small\
  \setlength{\tabcolsep}{5pt}
  \begin{tabular}{ c | c | c | c | c }
    \hline
    & \multicolumn{2}{c|}{Parameters} & Cost & $\Delta(\%)$ \\ \hline
    \multirow{3}{*}{\makecell{Network architecture}}
      & \multicolumn{2}{c|}{(32, 64)} & $1.1509 \times 10^4$  & 1.32\%\\
      & \multicolumn{2}{c|}{(512, 1024)} & $1.1439 \times 10^4$ & 0.70\% \\
      & \multicolumn{2}{c|}{(128, 256, 128)} & $1.1442 \times 10^4$ & 0.73\% \\
    \hline
    \multirow{3}{*}{\makecell{Dimensions  $(n_z,n_v)$}} 
      & \multicolumn{2}{c|}{($n_z=50, n_v=4$)} & $1.1272\times 10^4$  & -0.77\%\\
      & \multicolumn{2}{c|}{($n_z=80,n_v=4$)} & $1.1351 \times 10^4$ & -0.07\% \\
      & \multicolumn{2}{c|}{($n_z=60,n_v=8$)} & $1.1453 \times 10^4$ & 0.83\% \\
    \hline
    \multirow{2}{*}{\makecell{Training data size}}
      & \multicolumn{2}{c|}{1050}  & $1.1370 \times 10^4$ & 0.10\% \\
      & \multicolumn{2}{c|}{1750} & $1.1447 \times 10^4$ & 0.77\%\\
    \hline
    \multirow{4}{*}{Regularization} 
      & \multirow{2}{*}{$\beta_z$} & $4\times 10^{10}$ & $1.1340 \times 10^4$ & -0.17\% \\
      & & $6\times 10^{10}$ & $1.1526 \times 10^4$ & 1.47\% \\
    \cline{2-5}
      & \multirow{2}{*}{$\beta_g$} & $5\times 10^{-8}$ & $1.1313 \times 10^4$  & -0.40\%\\
      & & $5\times 10^{-9}$ & $1.1424 \times 10^4$ & 0.57\%\\
    \hline
  \end{tabular}
\end{table}

\subsection{Hyperparameter sensitivity analysis}
We further investigate the impact of hyperparameters on the performance of the proposed regularized DeeEPC for the wastewater treatment process. Specifically, we examine how the neural network architecture, the dimensions of the lifted variables, the size of the training dataset, and the regularization parameters influence the economic cost. The average economic costs under dry weather for different configurations are shown in Table~\ref{table:sensiticity}. $\Delta (\%)$ in Table~\ref{table:sensiticity} is the percentage differences between the corresponding configurations and the Regularized DeeEPC baseline introduced in Section~\ref{section:settings}, computed as $\Delta(\%)=\frac{C_{\text{config}}-C_{\text{base}}}{C_{\text{base}}}\times 100\%$, where $C_{\text{config}}$ is the average economic cost for a specific setting, $C_{\text{base}}$ is the average economic cost for the Regularized DeeEPC baseline. Simulation results indicate that the proposed method is relatively robust to variations in these hyperparameters. The average economic cost deviates from that of the regularized DeeEPC baseline by less than 1.5\% across all the tested configurations. Moreover, in most configurations, the proposed method achieves lower economic costs than the other control methods. These results suggest that the proposed method requires only limited effort in hyperparameter tuning.

\section{Case study on shipboard carbon capture system}
\begin{figure*}[t!]
    \centering
    \centering
    \includegraphics[width=0.89\textwidth]{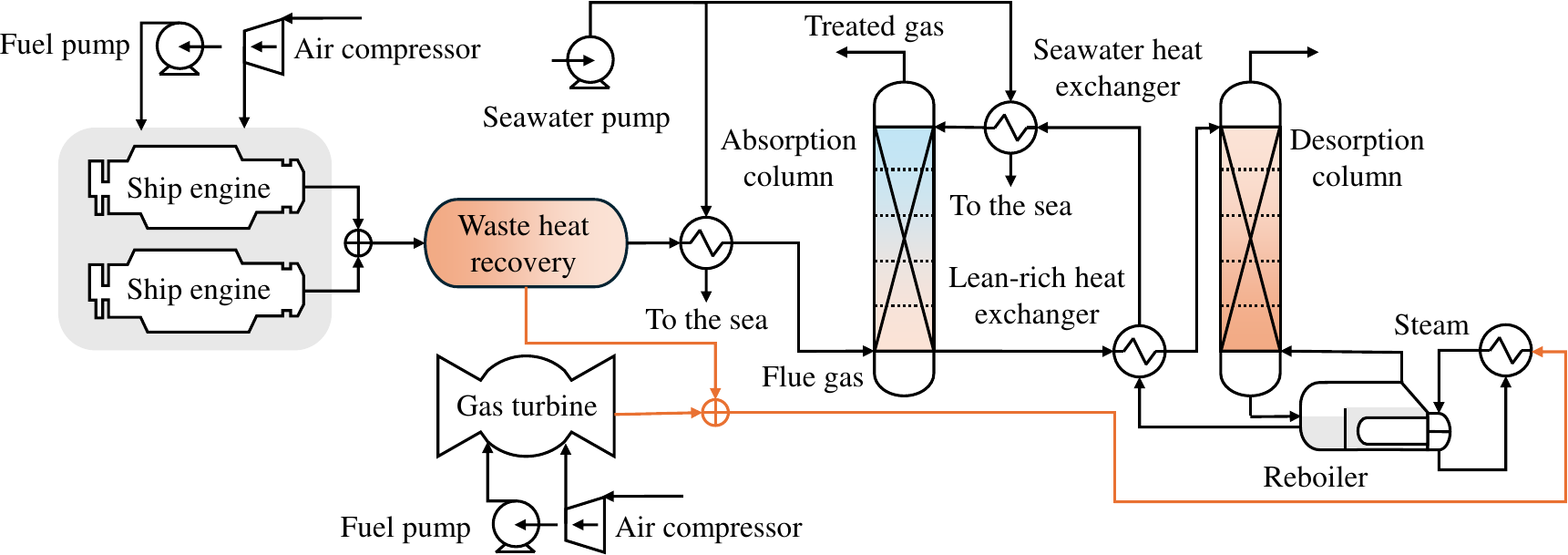}
    \caption{A schematic of the shipboard carbon capture system~\cite{zhang2025machine}.}
    \label{fig:spcc}
\end{figure*}

\subsection{Process description}

A shipboard carbon capture process involves installing a facility on a ship that connects to the ship engine(s) to capture carbon dioxide from exhaust gases, thereby reducing greenhouse gas emissions from maritime operations~\cite{sridhar2024technoeconomic,awoyomi2019co2,vo2024advanced}. In this section, we apply the proposed method to a shipboard solvent-based post-combustion carbon capture process, which is presented in~\cite{zhang2025machine}. A schematic of this carbon capture process is given in Figure~\ref{fig:spcc}. The ship engine system comprises two diesel engines and a diesel gas turbine. The two diesel engines provide propulsion power and generate hot flue gases, whose heat is recovered in a waste heat recovery system to supply energy to the reboiler. The diesel gas turbine combusts additional fuel to deliver supplementary heat to the reboiler. The solvent-based carbon capture plant is coupled with the ship engine system. In the absorption column, most of the CO$_2$ in the flue gases is captured. In the desorption column, the CO$_2$ absorbed in the solvent is released through thermal regeneration, using the heat provided by the reboiler. The solvent is subsequently cooled in heat exchangers and recycled back to the absorption column. More detailed information of the system is referred to~\cite{zhang2025machine}.


The dynamics of this process are characterized by a set of differential-algebraic equations with 103 differential state variables, 7 algebraic state variables, and 1 disturbance (i.e., the engine load $\varphi_E$). The dynamic model and a detailed description of the equations can be found in~\cite{zhang2025machine}. The dynamic model of this carbon capture process, as presented in~\cite{zhang2025machine}, is used as a dynamic process simulator. The control input vector contains the liquid solvent flow rate $F_L$ in $(\text{m}^3/\text{s})$, the fuel consumption rate of the diesel gas turbine $\tilde{F}_{fuel}$ in (kg/s), and the flow rate of the seawater $F_{sw}$ in $(\text{m}^3/\text{s})$, that is, $u = [F_L,\tilde{F}_{fuel}, F_{sw}]^\top$. The system output vector includes the flow rate of $\text{CO}_2$ in the treated gas $\tilde{F}_{\text{CO}_2}$ in (kg/s), the temperature of reboiler $T_{reb}$ in (K), and the carbon capture rate $p_{\text{CO}_2}$ (in $\%$), that is, $y = [\tilde{F}_{\text{CO}_2},T_{reb}, p_{\text{CO}_2}]^\top$. Specifically, the $\tilde{F}_{\text{CO}_2}$ is computed as $\tilde{F}_{\text{CO}_2}=r_{\text{CO}_2} C_{G,{\text{CO}_2}}^1 F_G$, where $r_{\text{CO}_2}$ is the molar mass of $\text{CO}_2$, $C_{G,{\text{CO}_2}}^1$ is the concentration of $\text{CO}_2$ in the first layer of the absorption column, and $F_G$ is the gas flow rate; the capture rate is calculated as $ p_{\text{CO}_2}= (\tilde{F}_{flue,\text{CO}_2} - \tilde{F}_{\text{CO}_2})/\tilde{F}_{flue,\text{CO}_2}$, where $\tilde{F}_{flue,\text{CO}_2}$ represents the mass flow rate of $\text{CO}_2$ in the flue gas.

The economic cost function accounts for the $\text{CO}_2$ emission tax and energy consumption~\cite{zhang2025machine,han2025deep}:
\begin{equation}\label{spcc_e}
    \ell_e(y_k,u_k) = \beta_1 \text{max}\big((y_{k,(1)}  - y_{limit}),0\big) + \beta_2 u_{k,(2)}
\end{equation}
where $\beta_1$ is the carbon tax (\$/kg); $\beta_2$ is the fuel price (\$/kg), $y_{limit}$ is the carbon release threshold (kg/s).

\subsection{Control methods and parameters}
We examine the scenario where the first-principles model is unknown. The performances of two data-driven control approaches are assessed, including the regularized DeeEPC and a learning-based Koopman EMPC proposed in~\cite{han2025deep}, referred to as deep neural Koopman EMPC in the remainder of this section. 

The hyperparameters of the proposed regularized DeeEPC are $T = 10^3$, $T_{ini}=2$, $N_p=2$, $n_z = 32$, and $n_v = 4$. $n_z$ and $n_v$ are selected to optimize the training performance. The regularization parameters are $\beta_z = 1\times 10^{4}$ and $\beta_g = 1$, respectively. The hyperparameters for training the proposed controller are made the same as those in Table~\ref{table:trainingparameter}. The regularized DeeEPC is initialized by applying PID control~\cite{seborg2016process} that tracks the set-point $\tilde{F}_{\text{CO}_2} = 2193.46$ (kg/h) and $T_{reb} = 391.73$ (K).
In the deep neural Koopman EMPC designed following~\cite{han2025deep}, the Koopman model is built in a lifted space with 40 latent state variables, and the model is trained over 100 epochs. The length of the control horizon for deep neural Koopman EMPC is set to 10 for online control implementation. Details are as provided in~\cite{han2025deep}.

\begin{table}[t]
  \renewcommand\arraystretch{1.25}
  \caption{Load ranges for three ship operational conditions as adapted from~\cite{han2025deep}.}\label{table:load_range}\vspace{2mm}
  \centering\small
    \begin{tabular}{ c c c}
      \toprule
       & Range of engine load $\varphi_E$\\
      \midrule
      Maneuvering & $70\% - 100\%$\\
      Slow steaming & $40\% - 70\%$\\
      Low engine load &  $20\% - 40\%$\\ 
      \bottomrule
    \end{tabular}
\end{table}

\begin{table}[t]
  \renewcommand\arraystretch{1.25}
  \caption{Control input limits for the shipboard carbon capture process.}\label{table:inputbound_spcc}\vspace{2mm}
  \centering\small
    \begin{tabular}{ c c c}
      \toprule
       & lower bound & upper bound \\
      \midrule
      $F_L$ $(\text{m}^3/\text{s})$ & 30  & 150\\
      $\tilde{F}_{fuel}$ (kg/h) & 2500  & 5000 \\
      $F_{sw}$ $(\text{m}^3/\text{s})$ & 20 & 90\\
      \bottomrule
    \end{tabular}
\end{table}

\begin{figure}[!t]
    \centering
    \includegraphics[width=0.48\textwidth]{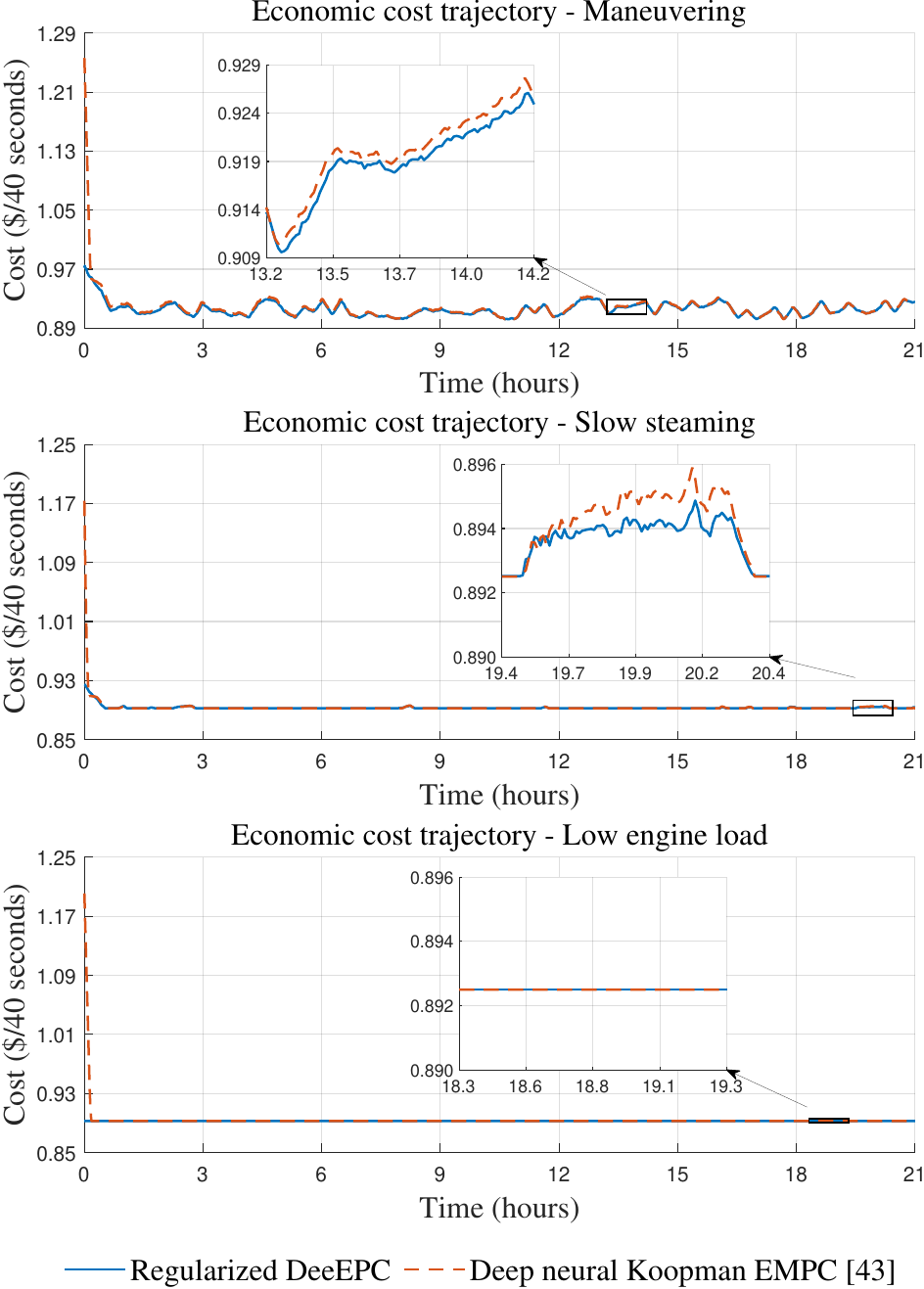}
    \caption{The trajectories of the operational economic costs under the three ship operational conditions.}
    \label{fig:spcc_cost}
\end{figure}

\subsection{Data generation}
Open-loop simulations with a sampling period $\Delta = 40$ s are conducted to collect data for training both the proposed regularized DeeEPC controller and the deep neural Koopman EMPC controller~\cite{han2025deep}, during which the engine load fluctuates between 20\% and 100\% randomly, with each steady load period lasting a randomly selected interval of 400–1200 s. 
The trained controllers are evaluated under three operational conditions, including maneuvering, slow steaming, and low engine load to assess the robustness of the controllers in the presence of unseen disturbances. 
The engine load ranges corresponding to the three operating conditions are presented in Table~\ref{table:load_range}. Three ship operational conditions, adapted from~\cite{han2025deep}, are considered. 
The constraints on the three control input variables are shown in Table~\ref{table:inputbound_spcc}. Hard output constraints are imposed on the reboiler temperature $T_{reb}$, that is, $T_{reb}\in [385.15\ (\text{K}), 393.15\ (\text{K})]$.


A dataset that contains $4.5\times 10^3$ input and output data is generated for training of the proposed regularized DeeEPC, among which $10^3$ input and output samples are used to formulate the Hankel matrices. The training of the deep neural Koopman EMPC~\cite{han2025deep} requires more data to establish an accurate Koopman model compared to the proposed method. $5.4\times 10^4$ samples are used to train the Koopman EMPC controller. In addition, the method in~\cite{han2025deep} requires online measurements of more state variables, including the liquid and gas temperatures at each layer of the absorption and desorption columns. Measurements of 20 states and the disturbance are assumed to be available and used to establish the Koopman model.


\begin{table}[!htb]
\centering
\caption{Average hourly economic costs (\$/h) under three ship operational conditions.}\small\vspace{2mm}\label{table:spcc_avgcost}
\renewcommand\arraystretch{1.25}
\centering \small
\begin{tabular}{>{\centering\arraybackslash}p{2.3cm} 
                >{\centering\arraybackslash}p{2cm} 
                >{\centering\arraybackslash}p{3cm}}    
\toprule
& Regularized DeeEPC & Deep neural Koopman EMPC~\cite{han2025deep} \\
\midrule
Maneuvering & $82.4478$ &   $82.6651$\\
Slow steaming & $80.3769$ &  $80.4365$ \\
Low engine load & $80.3250$ &   $80.4461$ \\
\bottomrule
\end{tabular}
\end{table}

\subsection{Control performance}

The trajectories of the economic costs given by the proposed method and the deep neural Koopman EMPC in~\cite{han2025deep} are given in Figure~\ref{fig:spcc_cost}. The average hourly economic costs of the two methods are shown in Table~\ref{table:spcc_avgcost}.
The economic costs for the simulated shipboard carbon capture system presented in Figure~\ref{fig:spcc_cost}  and Table~\ref{table:spcc_avgcost} are evaluated using the economic cost function defined in~(\ref{spcc_e}).
As shown in Figure~\ref{fig:spcc_cost} and Table~\ref{table:spcc_avgcost}, the proposed method achieves lower economic costs than the baseline~\cite{han2025deep} under all three operational conditions. The proposed approach reduces the average economic cost by 0.07\%–0.26\% compared with the deep neural Koopman EMPC~\cite{han2025deep}.
We note that the amount of released $\text{CO}_2$ increases with ship engine load. Consequently, the average economic cost rises at higher ship engine loads due to the greater energy required to capture the increased $\text{CO}_2$ emissions.

\section{Conclusion}
In this work, a convex data-enabled economic predictive control (DeeEPC) approach was proposed for a class of general nonlinear systems. This approach involves two lifting functions, in the form of two neural networks, that generate mapped output and mapped input respectively, which are used to construct a quadratic function to approximate the economic cost of the underlying system. Key system outputs are reconstructed from the mapped output, and hard output constraints on these outputs are incorporated into the formulated convex optimization problem for online control. We conducted theoretical analysis to establish the suitability of using the proposed framework in handling nonlinear dynamics and approximating the associated economic cost function. The training process was presented, following which the parameters of the two neural networks (accounting for output mapping and input mapping), the quadratic approximation of the economic cost, and the output reconstruction matrix can be learned from open-loop data from the underlying nonlinear system. We validated the proposed method on two simulated large-scale industrial processes, including a water treatment process and a ship-based carbon capture process. From the results, hard constraints on key process outputs are satisfied. Comparisons showed that the proposed method achieves better economic performance, as compared to benchmark control designs.  


\bibliographystyle{elsarticle-num}
\bibliography{arxiv}

@article{koopman1931hamiltonian,
  title={Hamiltonian systems and transformation in {Hilbert} space},
  author={Koopman, Bernard O},
  journal={Proceedings of the National Academy of Sciences},
  volume={17},
  number={5},
  pages={315--318},
  year={1931},
  publisher={National Acad Sciences}
}

@article{1428856,
  title={A note on persistency of excitation},
  author={Willems, Jan C and Rapisarda, Paolo and Markovsky, Ivan and De Moor, Bart LM},
  journal={Systems \& Control Letters},
  volume={54},
  number={4},
  pages={325--329},
  year={2005},
  publisher={Elsevier}
}

@article{van2020willems,
  title={Willems’ fundamental lemma for state-space systems and its extension to multiple datasets},
  author={Van Waarde, Henk J and De Persis, Claudio and Camlibel, M Kanat and Tesi, Pietro},
  journal={IEEE Control Systems Letters},
  volume={4},
  number={3},
  pages={602--607},
  year={2020},
  publisher={IEEE}
}

@inproceedings{coulson2019data,
  title={Data-enabled predictive control: In the shallows of the DeePC},
  author={Coulson, Jeremy and Lygeros, John and D{\"o}rfler, Florian},
  booktitle={European Control Conference},
  pages={307--312},
  year={2019}
}

@article{korda2018linear,
  title={Linear predictors for nonlinear dynamical systems: {Koopman} operator meets model predictive control},
  author={Korda, Milan and Mezi{\'c}, Igor},
  journal={Automatica},
  volume={93},
  pages={149--160},
  year={2018},
  publisher={Elsevier}
}

@article{li2024machine,
  title={Machine learning-based input-augmented {Koopman} modeling and predictive control of nonlinear processes},
  author={Li, Zhaoyang and Han, Minghao and Vo, Dat-Nguyen and Yin, Xunyuan},
  journal={Computers \& Chemical Engineering},
  volume={191},
  pages={108854},
  year={2024},
  publisher={Elsevier}
}

@article{han2024efficient,
  title={Efficient economic model predictive control of water treatment process with learning-based {Koopman} operator},
  author={Han, Minghao and Yao, Jingshi and Law, Adrian Wing-Keung and Yin, Xunyuan},
  journal={Control Engineering Practice},
  volume={149},
  pages={105975},
  year={2024},
  publisher={Elsevier}
}

@book{kashin2005orthogonal,
  title={Orthogonal series},
  author={Kashin, Boris Sergeevic and Saakyan, Artur Artushovich},
  volume={75},
  year={2005},
  publisher={American Mathematical Soc.}
}

@book{bauer2001measure,
  title={Measure and integration theory},
  author={Bauer, Heinz},
  volume={26},
  year={2001},
  publisher={Walter de Gruyter}
}

@book{steele2004cauchy,
  title={The Cauchy-Schwarz master class: an introduction to the art of mathematical inequalities},
  author={J. Michael Steele},
  year={2004},
  publisher={Cambridge University Press}
}

@book{giaquinta2010mathematical,
  title={Mathematical analysis: An introduction to functions of several variables},
  author={Giaquinta, Mariano and Modica, Giuseppe},
  year={2010},
  publisher={Springer Science \& Business Media}
}

@incollection{muscat2024hilbert,
  title={Hilbert Spaces},
  author={Muscat, Joseph},
  booktitle={Functional Analysis: An Introduction to Metric Spaces, Hilbert Spaces, and Banach Algebras},
  pages={197--247},
  year={2024},
  publisher={Springer}
}

@article{cabrera1995hilbert,
  title={Hilbert modules revisited: Orthonormal bases and Hilbert-Schmidt operators},
  author={Cabrera, M and Martinez, J and Rodriguez, A},
  journal={Glasgow Mathematical Journal},
  volume={37},
  number={1},
  pages={45--54},
  year={1995},
  publisher={Cambridge University Press}
}

@article{bibby1974axiomatisations,
  title={Axiomatisations of the average and a further generalisation of monotonic sequences},
  author={Bibby, John},
  journal={Glasgow Mathematical Journal},
  volume={15},
  number={1},
  pages={63--65},
  year={1974},
  publisher={Cambridge University Press}
}

@article{zhang2023dimension,
  title={Dimension reduction for efficient data-enabled predictive control},
  author={Zhang, Kaixiang and Zheng, Yang and Shang, Chao and Li, Zhaojian},
  journal={IEEE Control Systems Letters},
  year={2023},
  publisher={IEEE}
}

@article{alex2008benchmark,
  title={Benchmark simulation model no. 1 (BSM1)},
  author={Alex, Jens and Benedetti, Lorenzo and Copp, JB and Gernaey, KV and Jeppsson, Ulf and Nopens, Ingmar and Pons, MN and Rieger, Leiv and Rosen, Christian and Steyer, JP and others},
  journal={Report by the IWA Taskgroup on benchmarking of control strategies for WWTPs},
  volume={1},
  year={2008},
  publisher={Department of Industrial Electrical Engineering and Automation, Lund~…}
}

@article{zeng2015economic,
  title={Economic model predictive control of wastewater treatment processes},
  author={Zeng, Jing and Liu, Jinfeng},
  journal={Industrial \& Engineering Chemistry Research},
  volume={54},
  number={21},
  pages={5710--5721},
  year={2015},
  publisher={ACS Publications}
}

@article{kingma2014adam,
  title={Adam: A method for stochastic optimization},
  author={Kingma, Diederik P and Ba, Jimmy},
  journal={arXiv preprint arXiv:1412.6980},
  year={2014}
}

@misc{waterweb,
    title = {International {Water} {Association}},
    howpublished = {\url{http://www.benchmarkwwtp.org}},
    key = {IWA}
}

@article{wachter2006implementation,
  title={On the implementation of an interior-point filter line-search algorithm for large-scale nonlinear programming},
  author={W{\"a}chter, Andreas and Biegler, Lorenz T},
  journal={Mathematical programming},
  volume={106},
  pages={25--57},
  year={2006},
  publisher={Springer}
}

@article{zhang2025machine,
  title={Machine learning-based hybrid dynamic modeling and economic predictive control of carbon capture process for ship decarbonization},
  author={Zhang, Xuewen and Huang, Kuniadi Wandy and Vo, Dat-Nguyen and Han, Minghao and Decardi-Nelson, Benjamin and Yin, Xunyuan},
  journal={arXiv preprint arXiv:2502.05833},
  year={2025}
}

@article{han2025deep,
  title={Deep Neural {Koopman} Operator-Based Economic Model Predictive Control of Shipboard Carbon Capture System},
  author={Han, Minghao and Yin, Xunyuan},
  journal={IEEE Transactions on Control Systems Technology},
  year={2025},
  publisher={IEEE}
}

@article{brunton2021modern,
  title={Modern {Koopman} theory for dynamical systems},
  author={Brunton, Steven L and Budi{\v{s}}i{\'c}, Marko and Kaiser, Eurika and Kutz, J Nathan},
  journal={arXiv preprint arXiv:2102.12086},
  year={2021}
}

@article{linares2019koopman,
  title={{Koopman} operator theory applied to the motion of satellites},
  author={Linares, Richard},
  journal={Advances in the Astronautical Sciences},
  volume={171},
  year={2019}
}

@article{servadio2023koopman,
  title={{Koopman}-operator control optimization for relative motion in space},
  author={Servadio, Simone and Armellin, Roberto and Linares, Richard},
  journal={Journal of Guidance, Control, and Dynamics},
  volume={46},
  number={11},
  pages={2121--2132},
  year={2023},
  publisher={American Institute of Aeronautics and Astronautics}
}

@book{seborg2016process,
  title={Process dynamics and control},
  author={Seborg, Dale E and Edgar, Thomas F and Mellichamp, Duncan A and Doyle III, Francis J},
  year={2016},
  publisher={John Wiley \& Sons}
}

@article{yan2025economic,
author = {Mingxue Yan and Xuewen Zhang and Kaixiang Zhang and Zhaojian Li and Xunyuan Yin},
title = {Economic data-enabled predictive control using machine learning},
journal = {IFAC-PapersOnLine},
volume = {59},
number = {6},
pages = {25-30},
year = {2025},
note = {14th IFAC Symposium on Dynamics and Control of Process Systems, including Biosystems (DYCOPS 2025)}
}

@inproceedings{han2020deep,
  title={Deep learning of {Koopman} representation for control},
  author={Han, Yiqiang and Hao, Wenjian and Vaidya, Umesh},
  booktitle={IEEE Conference on Decision and Control},
  pages={1890--1895},
  year={2020},
}

@article{lusch2018deep,
  title={Deep learning for universal linear embeddings of nonlinear dynamics},
  author={Lusch, Bethany and Kutz, J Nathan and Brunton, Steven L},
  journal={Nature Communications},
  volume={9},
  number={1},
  pages={4950},
  year={2018},
  publisher={Nature Publishing Group UK London}
}

@article{shi2022deep,
  title={Deep {Koopman} operator with control for nonlinear systems},
  author={Shi, Haojie and Meng, Max Q-H},
  journal={IEEE Robotics and Automation Letters},
  volume={7},
  number={3},
  pages={7700--7707},
  year={2022},
  publisher={IEEE}
}

@book{henson1997nonlinear,
  title={Nonlinear process control},
  author={Henson, Michael A and Seborg, Dale E},
  year={1997},
  publisher={Prentice Hall PTR Upper Saddle River, New Jersey}
}

@article{schwenzer2021review,
  title={Review on model predictive control: An engineering perspective},
  author={Schwenzer, Max and Ay, Muzaffer and Bergs, Thomas and Abel, Dirk},
  journal={The International Journal of Advanced Manufacturing Technology},
  volume={117},
  number={5},
  pages={1327--1349},
  year={2021},
  publisher={Springer}
}

@article{rawlings2000tutorial,
  title={Tutorial overview of model predictive control},
  author={Rawlings, James B},
  journal={IEEE Control Systems Magazine},
  volume={20},
  number={3},
  pages={38--52},
  year={2000},
  publisher={IEEE}
}

@book{grune2017nonlinear,
  title={Nonlinear model predictive control},
  author={Gr{\"u}ne, Lars and Pannek, J{\"u}rgen and Gr{\"u}ne, Lars and Pannek, J{\"u}rgen},
  year={2017},
  publisher={Springer}
}

@article{ellis2014tutorial,
  title={A tutorial review of economic model predictive control methods},
  author={Ellis, Matthew and Durand, Helen and Christofides, Panagiotis D},
  journal={Journal of Process Control},
  volume={24},
  number={8},
  pages={1156--1178},
  year={2014},
  publisher={Elsevier}
}

@article{ellis2017economic,
  title={Economic model predictive control},
  author={Ellis, Matthew and Liu, Jinfeng and Christofides, Panagiotis D},
  journal={Springer},
  volume={5},
  number={7},
  pages={65},
  year={2017},
  publisher={Springer}
}

@article{morari2012nonlinear,
  title={Nonlinear offset-free model predictive control},
  author={Morari, Manfred and Maeder, Urban},
  journal={Automatica},
  volume={48},
  number={9},
  pages={2059--2067},
  year={2012},
  publisher={Elsevier}
}

@article{mayne2011tube,
  title={Tube-based robust nonlinear model predictive control},
  author={Mayne, David Q and Kerrigan, Erric C and Van Wyk, EJ and Falugi, Paola},
  journal={International journal of robust and nonlinear control},
  volume={21},
  number={11},
  pages={1341--1353},
  year={2011},
  publisher={Wiley Online Library}
}

@inproceedings{arbabi2018data,
  title={A data-driven {Koopman} model predictive control framework for nonlinear partial differential equations},
  author={Arbabi, Hassan and Korda, Milan and Mezi{\'c}, Igor},
  booktitle={IEEE Conference on Decision and Control},
  pages={6409--6414},
  year={2018},
}

@article{yan2025self,
  title={Self-tuning moving horizon estimation of nonlinear systems via physics-informed machine learning {K}oopman modeling},
  author={Yan, Mingxue and Han, Minghao and Law, Adrian Wing-Keung and Yin, Xunyuan},
  journal={AIChE Journal},
  volume={71},
  number={2},
  pages={e18649},
  year={2025},
  publisher={Wiley Online Library}
}

@inproceedings{yang2015data,
  title={A data-driven predictive controller design based on reduced Hankel matrix},
  author={Yang, Hua and Li, Shaoyuan},
  booktitle={Asian Control Conference},
  pages={1--7},
  year={2015}
}

@article{zhang2025deep,
  title={Deep {D}ee{PC}: Data-enabled predictive control with low or no online optimization using deep learning},
  author={Zhang, Xuewen and Zhang, Kaixiang and Li, Zhaojian and Yin, Xunyuan},
  journal={AIChE Journal},
  volume={71},
  number={3},
  pages={e18644},
  year={2025},
  publisher={Wiley Online Library}
}

@article{shang2024willems,
  title={Willems’ fundamental lemma for nonlinear systems with {Koopman} linear embedding},
  author={Shang, Xu and Cort{\'e}s, Jorge and Zheng, Yang},
  journal={IEEE Control Systems Letters},
  year={2024},
  publisher={IEEE}
}

@article{xie2023linear,
  title={Linear Data-Driven Economic {MPC} with Generalized Terminal Constraint},
  author={Xie, Yifan and Berberich, Julian and Allg{\"o}wer, Frank},
  journal={IFAC-PapersOnLine},
  volume={56},
  number={2},
  pages={5512--5517},
  year={2023},
  publisher={Elsevier}
}

@article{budivsic2012applied,
  title={Applied {Koopmanism}},
  author={Budi{\v{s}}i{\'c}, Marko and Mohr, Ryan and Mezi{\'c}, Igor},
  journal={Chaos: An Interdisciplinary Journal of Nonlinear Science},
  volume={22},
  number={4},
  year={2012},
  publisher={AIP Publishing}
}

@article{faulwasser2018economic,
  title={Economic nonlinear model predictive control},
  author={Faulwasser, Timm and Gr{\"u}ne, Lars and M{\"u}ller, Matthias A and others},
  journal={Foundations and Trends in Systems and Control},
  volume={5},
  number={1},
  pages={1--98},
  year={2018},
  publisher={Now Publishers, Inc.}
}

@inproceedings{rawlings2012fundamentals,
  title={Fundamentals of economic model predictive control},
  author={Rawlings, James B and Angeli, David and Bates, Cuyler N},
  booktitle={IEEE Conference on Decision and Control},
  pages={3851--3861},
  year={2012}
}

@article{tang2025big,
  title={Big Data-Driven Control of Nonlinear Processes Through Dynamic Latent Variables Using an Autoencoder},
  author={Tang, Jun Wen and Yan, Yitao and Bao, Jie and Huang, Biao},
  journal={IEEE Transactions on Cybernetics},
  volume={55},
  number={5},
  pages={2411--2423},
  year={2025},
}

@article{WR2025,
  title={Economic zone data-enabled predictive control for connected open water systems},
  author={Chen, Xiaoqiao and Zhang, Xuewen and Han, Minghao and Law, Adrian Wing-Keung and Yin, Xunyuan},
  journal={Water Research},
  volume={291},
  pages={125181},
  year={2026},
  publisher={Elsevier}
}

@article{pan2022stochastic,
  title={On a stochastic fundamental lemma and its use for data-driven optimal control},
  author={Pan, Guanru and Ou, Ruchuan and Faulwasser, Timm},
  journal={IEEE Transactions on Automatic Control},
  volume={68},
  number={10},
  pages={5922--5937},
  year={2022},
  publisher={IEEE}
}

@article{heidarinejad2012economic,
  title={Economic model predictive control of nonlinear process systems using {Lyapunov} techniques},
  author={Heidarinejad, Mohsen and Liu, Jinfeng and Christofides, Panagiotis D},
  journal={AIChE Journal},
  volume={58},
  number={3},
  pages={855--870},
  year={2012},
  publisher={Wiley Online Library}
}

@article{wu2020economic,
  title={{Economic MPC} of nonlinear processes via recurrent neural networks using structural process knowledge},
  author={Wu, Zhe and Rincon, David and Park, Michael and Christofides, Panagiotis D},
  journal={IFAC-PapersOnLine},
  volume={53},
  number={2},
  pages={11607--11613},
  year={2020},
  publisher={Elsevier}
}

@article{lao2014economic,
  title={Economic model predictive control of transport-reaction processes},
  author={Lao, Liangfeng and Ellis, Matthew and Christofides, Panagiotis D},
  journal={Industrial \& Engineering Chemistry Research},
  volume={53},
  number={18},
  pages={7382--7396},
  year={2014},
  publisher={ACS Publications}
}

@article{du2017real,
  title={Real-time microgrid economic dispatch based on model predictive control strategy},
  author={Du, Yan and Pei, Wei and Chen, Naishi and Ge, Xianjun and Xiao, Hao},
  journal={Journal of Modern Power Systems and Clean Energy},
  volume={5},
  number={5},
  pages={787--796},
  year={2017},
  publisher={SGEPRI}
}

@article{haque2020advanced,
  title={Advanced Process Control for Cost-Effective Glycol Loss Minimization in a Natural Gas Dehydration Plant under Upset Conditions},
  author={Haque, Md Emdadul and Palanki, Srinivas and Xu, Qiang},
  journal={Industrial \& Engineering Chemistry Research},
  volume={59},
  number={16},
  pages={7680--7692},
  year={2020},
  publisher={ACS Publications}
}

@article{sridhar2024technoeconomic,
  title={Technoeconomic evaluation of post-combustion carbon capture technologies on-board a medium range tanker},
  author={Sridhar, Preethi and Kumar, Anikesh and Manivannan, Sanjith and Farooq, Shamsuzzaman and Karimi, Iftekhar A},
  journal={Computers \& Chemical Engineering},
  volume={181},
  pages={108545},
  year={2024},
  publisher={Elsevier}
}

@article{awoyomi2019co2,
  title={{CO$_2$/SO$_2$} emission reduction in {CO}$_2$ shipping infrastructure},
  author={Awoyomi, Adeola and Patchigolla, Kumar and Anthony, Edward J},
  journal={International Journal of Greenhouse Gas Control},
  volume={88},
  pages={57--70},
  year={2019},
  publisher={Elsevier}
}

@article{vo2024advanced,
  title={Advanced designs and optimization for efficiently enhancing shipboard {CO}$_2$ capture},
  author={Vo, Dat-Nguyen and Zhang, Xuewen and Huang, Kuniadi Wandy and Yin, Xunyuan},
  journal={Industrial \& Engineering Chemistry Research},
  volume={63},
  number={48},
  pages={20963--20977},
  year={2024},
  publisher={ACS Publications}
}

@book{reed1980,
  title={Methods of modern mathematical physics. vol. 1. Functional analysis},
  author={Reed, Michael and Simon, Barry},
  year={1980},
  publisher={Academic New York}
}

@book{folland1999real,
  title={Real analysis: modern techniques and their applications},
  author={Folland, Gerald B},
  year={1999},
  publisher={John Wiley \& Sons}
}

@article{berberich2022linear,
  title={Linear tracking MPC for nonlinear systems—{Part II}: The data-driven case},
  author={Berberich, Julian and K{\"o}hler, Johannes and M{\"u}ller, Matthias A and Allg{\"o}wer, Frank},
  journal={IEEE Transactions on Automatic Control},
  volume={67},
  number={9},
  pages={4406--4421},
  year={2022},
  publisher={IEEE}
}

@article{huang2023robust,
  title={Robust data-enabled predictive control: Tractable formulations and performance guarantees},
  author={Huang, Linbin and Zhen, Jianzhe and Lygeros, John and D{\"o}rfler, Florian},
  journal={IEEE Transactions on Automatic Control},
  volume={68},
  number={5},
  pages={3163--3170},
  year={2023},
  publisher={IEEE}
}

@article{nonhoff2019economic,
  title={Economic model predictive control for snake robot locomotion},
  author={Nonhoff, Marko and K{\"o}hler, Philipp N and Kohl, Anna M and Pettersen, Kristin Y and Allg{\"o}wer, Frank},
  journal={IEEE Conference on Decision and Control},
  pages={8329--8334},
  year={2019},
  publisher={IEEE}
}

@article{huang2022economic,
  title={Economic model predictive control for multi-energy system considering hydrogen-thermal-electric dynamics and waste heat recovery of MW-level alkaline electrolyzer},
  author={Huang, Chunjun and Zong, Yi and You, Shi and Tr{\ae}holt, Chresten},
  journal={Energy Conversion and Management},
  volume={265},
  pages={115697},
  year={2022},
  publisher={Elsevier}
}

@incollection{floudas1995quadratic,
  title={Quadratic optimization},
  author={Floudas, Christodoulos A and Visweswaran, Viswanathan},
  booktitle={Handbook of global optimization},
  pages={217--269},
  year={1995},
  publisher={Springer}
}

@article{potra2000interior,
  title={Interior-point methods},
  author={Potra, Florian A and Wright, Stephen J},
  journal={Journal of computational and applied mathematics},
  volume={124},
  number={1-2},
  pages={281--302},
  year={2000},
  publisher={Elsevier}
}

@article{huang2021decentralized,
  title={Decentralized data-enabled predictive control for power system oscillation damping},
  author={Huang, Linbin and Coulson, Jeremy and Lygeros, John and D{\"o}rfler, Florian},
  journal={IEEE Transactions on Control Systems Technology},
  volume={30},
  number={3},
  pages={1065--1077},
  year={2021},
  publisher={IEEE}
}

@article{li2024physics,
  title={Physics-augmented data-enabled predictive control for eco-driving of mixed traffic considering diverse human behaviors},
  author={Li, Dongjun and Zhang, Kaixiang and Dong, Haoxuan and Wang, Qun and Li, Zhaojian and Song, Ziyou},
  journal={IEEE Transactions on Control Systems Technology},
  volume={32},
  number={4},
  pages={1479--1486},
  year={2024},
  publisher={IEEE}
}


\end{document}